\documentclass[sigconf, screen, anonymous=False]{acmart}
\renewcommand\footnotetextcopyrightpermission[1]{}
\PassOptionsToPackage{dvipsnames}{xcolor}
\PassOptionsToPackage{colorlinks=true,linkcolor=RubineRed,breaklinks=True,citecolor=RubineRed}{xcolor}

\usepackage{enumitem}
\setlist[itemize]{noitemsep, topsep=0pt}

\usepackage[most]{tcolorbox}
\usepackage{adjustbox}
\usepackage[linesnumbered,algo2e,ruled]{algorithm2e}
\usepackage{multirow}
\usepackage{booktabs,subcaption,dcolumn}
\usepackage{tikz}
\usepackage{enumitem}
\usepackage{tabularx}

\usepackage{pifont}
\usepackage{svg}
\usepackage{soul}
\usepackage{amsmath}
\usepackage{dirtytalk}
\usepackage{subcaption}
\usepackage{graphicx}
\usepackage{setspace}
\usepackage[font=small]{caption}
\usepackage{booktabs} 
\usepackage[titletoc,toc,title]{appendix}
\usepackage{svg}
\usepackage{amsfonts}
\usepackage{xurl} 

\usepackage{diagbox}
\usepackage{indentfirst}

\usepackage{amsmath}
\usepackage{amsfonts} 

\usepackage{colortbl}

\AtBeginDocument{%
  \providecommand\BibTeX{{%
    \normalfont B\kern-0.5em{\scshape i\kern-0.25em b}\kern-0.8em\TeX}}}

\setcopyright{none}

\acmDOI{xx.xxx/xxx_x}

\acmISBN{979-8-4007-0243-3/24/04}

\acmConference[SAC'24]{ACM SAC Conference}{April 8--April 12, 2024}{Avila, Spain}
\acmYear{2024}
\copyrightyear{2024}

\acmArticle{4}
\acmPrice{15.00}

\newcolumntype{?}{!{\vrule width 1.5pt}}

\newcommand{\textbox}[1]{
    \noindent\fbox{%
        \parbox{0.97\columnwidth}{%
            {#1}
        }%
    }
}

\newtcolorbox{cooltextbox}[1][]{%
    colback=black!5,
    colframe=black!5,
    notitle,
    sharp corners,
    borderline west={0pt}{0pt}{red!80!black},
    enhanced,
    breakable,
    left=0pt,
    right=0pt,
    top=0pt,
    bottom=0pt
    }

\newcommand\smamath[1]{{\small $#1$}}
\newcommand\smacal[1]{{\small $\mathcal{#1}$}}
\newcommand\smabb[1]{{\small $\mathbb{#1}$}}

\newcommand\scbb[1]{{\scriptsize $\mathbb{#1}$}}

\newcommand\scmath[1]{{\scriptsize $#1$}}

\newcommand\revision[1]{%
  \bgroup
  \hskip0pt\color{blue!80!black}%
  #1%
  \egroup
}

\begin{document}

\title{Understanding the Process of Data Labeling in Cybersecurity}

\author{Tobias Braun}
\orcid{0009-0001-5446-6469}
\affiliation{%
  \institution{University of Liechtenstein}
  \country{}
  }
\email{tobias.braun@uni.li}

\author{Irdin Pekaric}
\orcid{0000-0002-0706-3202}
\affiliation{%
  \institution{University of Liechtenstein}
  \country{}
  }
\email{irdin.pekaric@uni.li}

\author{Giovanni Apruzzese}
\orcid{0000-0002-6890-9611}
\affiliation{%
  \institution{University of Liechtenstein}
  \country{}
  }
\email{giovanni.apruzzese@uni.li}

\begin{abstract}

Many domains now leverage the benefits of Machine Learning~(ML), which promises solutions that can autonomously learn to solve complex tasks by training over some data. Unfortunately, in cyberthreat detection, high-quality data is hard to come by. Moreover, for some specific applications of ML, such data must be labeled by human operators. 
Many works ``assume'' that labeling is tough/challenging/costly in cyberthreat detection, thereby proposing solutions to address such a hurdle.
Yet, we found no work that specifically addresses the process of labeling \textit{from the viewpoint of ML security practitioners}. This is a problem: to this date, it is still mostly unknown how labeling is done in practice---thereby preventing one from pinpointing ``what is needed'' in the real world. 

In this paper, we take the first step to build a bridge between academic research and security practice in the context of data labeling. First, we reach out to five subject matter experts and carry out open interviews to identify pain points in their labeling routines. Then, by using our findings as a scaffold, we conduct a user study with 13 practitioners from large security companies, and ask detailed questions on subjects such as active learning, costs of labeling, and revision of labels. Finally, we perform proof-of-concept experiments addressing labeling-related aspects in cyberthreat detection that are sometimes overlooked in research. Altogether, our contributions and recommendations serve as a stepping stone to future endeavors aimed at improving the quality and robustness of ML-driven security systems. 
We release our resources.

\end{abstract}

%
%


\settopmatter{printfolios=true}

\keywords{Labeling, ML, Practitioners, User Study, Cyberthreat Detection}

\maketitle

\section{Introduction}
\label{sec:introduction}
\noindent
The never-ending advancements of Artificial Intelligence (AI) in research are in plain sight~\cite{kaur2022trustworthy,kaloudi2020ai}, and Machine Learning (ML) techniques are now becoming increasingly integrated also in operational information systems. Among the plethora of domains in which ML has found a real-world application (e.g.,~\cite{aghakhani2023venomave, shapira2023phantom}), the one of computer security -- and, in particular, \textbf{cyberthreat detection} -- stands out~\cite{apruzzese2023role}. On the one hand, by `training' ML models over some data, it is possible to develop ML-based systems that can mitigate the threat of zero-day attacks---which cannot be countered via conventional signature-based methods~\cite{dias2020go}. On the other hand, obtaining the data required to devise such data-driven solutions is challenging---especially from an organizational perspective~\cite{apruzzese2022sok}. 

Indeed, it is well-known that ``there is not such a thing as a foolproof system''~\cite{carlini2021poisoning}, therefore it is understandable that even ML-powered defenses may fail to detect all attacks. However, while some misclassifications may not raise serious security concerns, others may conceal signs of sophisticated attacks (e.g.,~\cite{lemay2018survey,ghafir2018detection}), which can lead to an entire organization becoming compromised~\cite{apruzzese2023role}. Simultaneously, security analysts are often overwhelmed by the excessive amount of false alarms that are raised by data-driven detectors~\cite{alahmadi202299}. The sheer reality is that the development of ML models ready for operational cybersecurity requires the collection of data points that pertain to the \textit{specific environment}\footnote{The requirement for environment-specific data is in \textbf{stark contrast} with many other applications of ML~\cite{apruzzese2022sok}. For instance, in visual object recognition, ``a cat will always be a cat'' and ``a dog will always be a dog''; in contrast, in cybersecurity, an IP address (or an URL) can be `benign' for one organization, and `malicious' for another one.} being monitored~\cite{apruzzese2022sok}. 

Such a peculiarity hence prevents a reliable `transfer' of ML models between different environments~\cite{apruzzese2022cross,de2023generalizing,catillo2022transferability}, which intrinsically hinders the advancement (both in research and practice) of ML for security applications.\footnote{The ML-based solutions proposed in many papers -- despite showing near-perfect accuracy -- have never seen the light of realistic deployment~\cite{apruzzese2023sok}. As a matter of fact, security practitioners see ML (and especially research papers~\cite{apruzzese2023sok}) with skepticism~\cite{de2019information}.} 
To give an idea, some security companies revealed~\cite{apruzzese2023role} that deployment of an ML-powered detector required almost one month of data collection (and extensive fine-tuning) done in their customers' network---which are operations that must be performed \textit{manually} and under the \textit{responsibility} of the security company. 
To make things worse, the process of ``obtaining a suitable training dataset'' may not only entail the `collection' of the data but also its `annotation': in other words, there is a need to associate each data-point to a given \textit{label} that is used during the training phase of the ML model to guide its learning~\cite{joyce2021framework}. Such a procedure -- required for \textit{supervised} ML methods -- necessitates a human who carefully assigns every sample in a dataset to its ground truth. 

Due to these reasons, in the security domain, it is now acknowledged that ``[data] labeling is expensive''~\cite{apruzzese2022sok,guerra2022datasets}, and abundant efforts have attempted to address this issue. For instance, many papers discuss ways to `optimize' the labeling process (e.g., by proposing active learning strategies~\cite{pendlebury2019tesseract}), or `decrease the cost' of labeling (e.g., by assigning coarse labels~\cite{van2022deepcase}); others seek to reduce the amount of `labeled instances' required to develop proficient ML models (e.g., few-shot learning~\cite{xu2020method}). Finally, some works (e.g.,~\cite{nisioti2018intrusion}) advise that `unsupervised' ML methods are more appropriate for cyberthreat detection, due to the absence of a labeling requirement~\cite{dias2020go}. 
However, despite a rich literature on this subject, we asked ourselves: ``How prohibitive is labeling in practice?''

\vspace{-3mm}
\begin{figure}[!htbp]
    \centering
    \includegraphics[width=0.85\columnwidth]{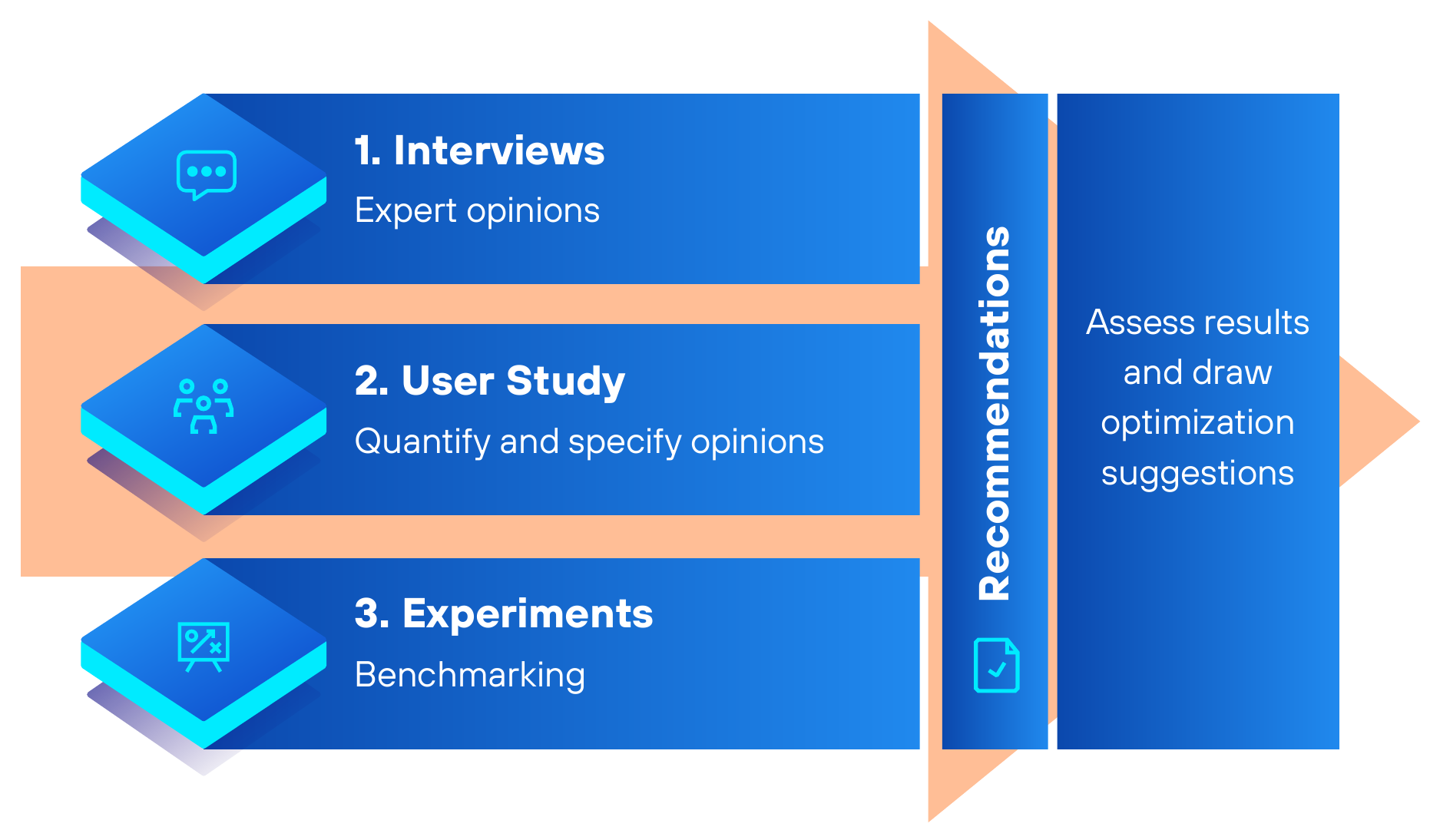}
    \vspace{-3mm}
    \caption{Overview (and contributions) of our paper.}
    \label{fig:contributions}
\end{figure}
\vspace{-3mm}

Perhaps surprisingly, we were unable to find any work that specifically investigated the \textit{labeling problem in itself}. To the best of our knowledge, most existing work assumed that labeling is costly, but no work studied how this process is carried out from an organizational perspective. The only evidence we were able to find was the 2016 paper by Miller et al.~\cite{miller2016reviewer}, which estimated that security companies may have a labeling budget of 80 samples per day. However, the security landscape has changed significantly since 2016~\cite{apruzzese2023role}, and (some) companies now do have labeling duties. Hence, in this paper, we seek to investigate the problem of data labeling from the viewpoint of security practitioners---with a focus on ML applications for cyberthreat detection.


\noindent
\textsc{\textbf{Contributions.}}
We seek to build a bridge between academic research and industrial practice in the context of data labeling for cybersecurity. To this purpose, after positioning our paper within existing literature (§\ref{sec:motivation}), we make the following contributions (Fig.~\ref{fig:contributions}):
\begin{itemize}[leftmargin=0.4cm]
    \item As a first step, we carry out \textbf{interviews with five security practitioners} having experience in ML development (§\ref{sec:interviews}). After qualitatively analyzing the responses, we highlight the challenges associated with data labeling, including pain points, costs, and time expenditures. Intriguingly, \textit{we found that even practitioners cannot estimate the costs of labeling}.
    
    \item By using our interviews as a scaffold, we conduct a \textbf{user study with security experts from 13 different companies} (§\ref{sec:user}). Our semi-structured questionnaire delves deeper into the subject of data labeling, and our quantitative analyses reveal how various security companies address this problem for real-world projects. Interestingly, \smamath{31\%} had \textit{never heard of the term ``active learning''}, and some stated that \textit{it leads to overconfidence}.

    \item To validate some of our previous findings, we perform \textbf{technical experiments focused on some labeling aspects overlooked in research}~(§\ref{sec:experiments}). Specifically, we showcase the importance of repeated evaluations in cases of scarce labeled data; the effects of mislabeling; and the pros-and-cons of active learning methods---\textit{which can reach plateaus with no practical benefits at all}.
\end{itemize}
After discussing our study (§\ref{sec:discussion}), we coalesce all of our novel findings into a set of \textbf{recommendations and takeaways} (§\ref{sec:conclusions}) useful to improve the state-of-the-art (for both research and practice). Finally, for scientific reproducibility and to pave the way for future work, we publicly release all our resources~\cite{ourRepo}. 
\section{Background and Motivation}
\label{sec:motivation}
\noindent
We summarize some well-known cybersecurity labeling strategies~(§\ref{ssec:background}), and then compare our paper against related work~(§\ref{ssec:related}).

\subsection{Labeling Strategies (in research)}
\label{ssec:background}
\noindent
The starting point of any application of machine learning is \textbf{data}, whose role is developing a given ML model, which will then be used to analyze `unseen' samples. In the context of cyberthreat detection, such \textit{training} data \textit{must} be provided with some reference information (i.e., a `label') used to distinguish benign from malicious samples. \footnote{Cyberthreat detection is $\perp$ to \textit{anomaly detection} (not all anomalies are ``a threat''~\cite{robertson2006using}).}
Obtaining such reference information, however, requires some human `supervision'.\footnote{Hence the name of \textit{supervised} ML techniques~\cite{sommer2010outside}.} 
Consequently, whenever labeling is required, it is common~\cite{apruzzese2022sok} to introduce some form of labeling \textit{budget}, \smacal{B}, used to associate each sample, \smamath{x}, of a given dataset, \smabb{D}, to its ground truth, \smamath{y}. From an organizational perspective, labeling can be seen as that process which uses \smacal{B} to obtain \smabb{L}, i.e., a \textit{labeled} dataset (having \smabb{L}\smamath{\subseteq}\smabb{D}) used to develop an ML model \smamath{M}.

Without loss of generality, we identify the following labeling approaches---discussed, adopted, and proposed in research (and sometimes associated to the term ``semi-supervised learning''~\cite{apruzzese2022sok}). 
\begin{itemize}[leftmargin=0.4cm]
    \item \textit{Random} (e.g.,~\cite{andresini2021gan}), i.e., the most naive form of labeling: After obtaining a given dataset, the annotator labels each sample without following any specific strategy, until the budget is depleted.

    \item \textit{Temporal} labeling (e.g.,~\cite{andresini2021insomnia}), whose goal is to label the samples according to their chronological occurrence (until the budget is depleted), to provide a ``temporally consistent'' testbed (useful to prevent temporal bias which may skew the results~\cite{arp2022and}).
    
    \item \textit{Crowdsourcing} (e.g.,~\cite{zhu2019tripartite}), in which the labeling efforts are delegated (by using the available budget) to third-parties. This is a common approach in computer vision~\cite{luccioni2023bugs}, which is receiving attention also in cybersecurity~\cite{hong2020phishing} (especially when the annotation does not require extensive domain expertise).

    \item \textit{Synthetic generation} (e.g.,~\cite{sharafaldin2018toward}), which entails the creation of specific samples whose ground truth is known ``a priori''.
    
    \item \textit{Active learning} (e.g.,~\cite{pendlebury2019tesseract}), which seeks to optimize the labeling procedure by ``suggesting'' specific samples to the annotator: The idea is to prioritize labeling of those samples that can maximize the learning of the ML model (see Appendix~\ref{app:active} for more details).
    
\end{itemize}
We also mention approaches typically denoted as \textit{self-supervised learning}, which revolve around having the ML model to (iteratively) learn on the (likely inaccurate) predictions that it makes when analyzing unlabeled data (e.g.,~\cite{silva2023self,liu2021self}).

We observe, however, that the practical effectiveness of all the abovementioned strategies is \textbf{questionable} or still unclear. For instance, random labeling is, by definition, inefficient (and is the source of experimental bias~\cite{apruzzese2022sok, arp2022and}). Temporal labeling requires accurate timestamps, which are not always available~\cite{apruzzese2023sok}. Crowdsourcing is reliant on the judgment (and `honesty'~\cite{matsuura2021careless}) of people who may not be at all interested in the performance of the resulting ML model~\cite{luccioni2023bugs}. Generating data synthetically can be economically viable (since the budget is virtually infinite), but it is difficult to do so in a realistic way (e.g., the generated data may represent threats that are well-known and for which there are already countermeasures~\cite{garcia2014empirical}) and recent research showed plenty of inaccuracies in some popular datasets~\cite{engelen2021troubleshooting}. Self-supervised learning has been recently shown to provide almost negligible benefits in cyberthreat detection~\cite{apruzzese2022sok}. Finally, while active learning (AL) has consistently proven to be advantageous~\cite{chen2020malware,gornitz2009active,li2017phishbox, pendlebury2019tesseract, apruzzese2022sok}, it is still unclear \textit{how to reliably use it in practice}: as we will show in this paper (§\ref{sec:user}), some practitioners are oblivious of the term ``active learning''.

\subsection{Labeling in Practice (related work)}
\label{ssec:related}
\noindent
Despite thousands of papers that focus on the interplay between ML and cybersecurity (see, e.g.,~\cite{apruzzese2023role, kaloudi2020ai, sarker2020cybersecurity} for literature surveys), we observed that no paper attempted to scrutinize \textbf{how labeling is done by security practitioners}. 

Indeed, we carried out an extensive review of existing cybersecurity literature, and we found that most works that seek to mitigate the problem of data-labeling simply acknowledge that ``labeling is costly in practice'', and then proceed to propose a solution that attempts to alleviate such costs. For example, in 2017, Li et al.~\cite{li2017phishbox} showed that using AL (w.r.t. random selection) allows a \textit{phishing detector} to converge faster (in terms of the number of labeled instances required) to its ideal maximum accuracy; a similar finding was made in 2020 by Chen et al.~\cite{chen2020malware} for \textit{malware classification}, and in 2021 by Zhang et al.~\cite{zhang2021network} for network intrusion detection. Yet, all these solutions have been assessed in a laboratory setting, and they did not undergo any form of validation by real practitioners---despite achieving substantial performance improvements.\footnote{Albeit, interestingly, a recent work~\cite{apruzzese2022sok} revealed that most comparisons presented shortcomings, thereby questioning whether prior work was effective even in research.} To the best of our knowledge, the few (recent) exceptions are the paper by Van Ede et al.~\cite{van2022deepcase}, encompassing authors from both industry and academia; and the study by Fredriksson et al.~\cite{fredriksson2020data}. However, the latter -- while providing insight from practitioners -- does not pertain to cybersecurity; whereas the former -- which proposes a coarse labeling strategy that is validated in a real SOC -- only accounts for the perspective of a single security company.

Put simply, scientific literature overlooks the \textit{real-world implications of data-labeling in cybersecurity}---which, to the best of our knowledge, are still unknown. This is a problem because \textbf{it prevents one} from determining: 
{\small (i)}~whether a given solution is truly applicable to a given context (i.e., does its adoption in practice yield some benefits?);
{\small (ii)}~which methods should be given more attention (i.e., by knowing which methods are used in practice, one can focus on improving such methods);
{\small (iii)}~the overall role of data-labeling in an operational workflow (i.e., do practitioners really care?).
Addressing any of these issues is, however, \textbf{challenging}--- especially from the perspective of a researcher. This is because doing so requires the researcher to go \textit{beyond the lab}, i.e., they must establish some form of collaboration with security professionals---whose practices are often kept hidden (both for their own companies' security, as well as for trade-secrets~\cite{hannah2005should}). For instance, receiving permission to test a given solution on a real security system may be unfeasible for researchers, whereas finding companies who are willing to disclose (parts of) their workflows is tough~\cite{meyer2022smartgrid}. In this paper, we aim to overcome all such challenges.

\begin{cooltextbox}
\textsc{\textbf{Problem.}}
Despite abundant works claiming that ``labeling is costly in cybersecurity'', there is no paper that attempted to investigate such a hurdle from the perspective of \textit{practitioners}. 
\end{cooltextbox}

\noindent
To shed light on the process of data labeling in operational cybersecurity, we reach out to security practitioners (through both expert interviews and user studies) and ask them to share some insights deriving from their daily routines.\footnote{\textbf{Ethics:} Our institutions know and approve this research. We follow the Menlo report.} We then carry out proof-of-concept experiments to validate some of the findings brought to light by our prior analyses. Such a twofold approach is \textbf{unusual in related literature}. Indeed, technical papers (e.g.,~\cite{gornitz2009active}) tend to overlook the perspective of practitioners; whereas papers that investigate the practitioners' viewpoint (e.g.,~\cite{fredriksson2020data}), do not perform any sort of validation---aside from not being focused on cybersecurity. 
\section{Expert Interviews}
\label{sec:interviews}

\noindent
Our first contribution are the findings of interviews with experts in the field of ML and cybersecurity. We describe the methodology~(§\ref{ssec:interviews_method}); then, we present~(§\ref{ssec:interviews_results}) and discuss~(§\ref{ssec:interviews_interpretation}) our results. 


\subsection{Method}
\label{ssec:interviews_method}

\noindent
The goal of these interviews was to identify pain points in the data labeling process, assess its costs and time factors, and gather insights through narrative and directed open-questions~\cite{weller2018open}.

\textbf{Participants.}
To identify suitable subject matter experts (SME), we reached out to over 40 companies with expertise in cybersecurity and ML. Companies actively involved in ML programming for cybersecurity applications were specifically targeted, rather than those solely using pre-existing ML solutions. Despite sending hundreds of emails, we found an agreement only with five SMEs, each representing a different company (located in Europe, and having >500 employees). Such SMEs agreed to share some information on their daily routines (due to NDA, we cannot reveal more information on our participants). All these difficulties (common in related studies, e.g.,~\cite{meyer2022smartgrid,alahmadi202299}) increase the value of our findings.

\textbf{Questions.}
 After reaching an agreement with our SME, two authors held various brainstorming sessions aimed at deriving a set of questions that would be used as a basis for the interviews. Specifically, we sought to frame open-ended questions that would facilitate a broad discussion, which allows for uncovering non-obvious issues---while accounting for potential NDA binding the interviewee. Eventually, we concocted eight questions, for which we also anticipated potential answers and prepared likely follow-up questions to delve deeper into specific issues based on the interviewee's responses. In particular, for each question, we predicted between one and five potential answers and defined between three to seven follow-up questions. This laid the foundation for gaining a comprehensive understanding of the practical implications, which formed the basis for subsequent endeavors such as user study and experiments aimed at addressing the identified issues within real-world contexts. Our generic set of questions is available in our public repository~\cite{ourRepo}, but we provide a summary in Table~\ref{table:question_topics}.

\begin{table}[!htbp]
    \centering
    \caption{Interview topics}
    \label{table:question_topics}
    \vspace{-3mm}
    \small
    \begin{tabular}{@{}|l|l|}
        \hline
        \textbf{Question No.} & \textbf{Category - Data Labeling} \\
        \hline
        \rowcolor[HTML]{DAE8FC} 
        1 & Description of the Process \\
        2 - 3 & Resource Requirements and Time Expenditure \\
        \rowcolor[HTML]{DAE8FC} 
        4 - 6 & Improvement Possibilities and Strategies \\
        7 - 8 & Future Predictions and Impact \\
        \hline
    \end{tabular}
\end{table}   
\vspace{-3mm}

\textbf{Conduction.}
The interviews were done (in English) by the same author, who reached out to each SME and agreed on a one-hour timeslot to have a remote interview. We did not prime our SMEs (i.e., we did not send them any questions beforehand), but they were informed that the interview would revolve around labeling practices. The interviews were not recorded, and the interviewer, after asking each question, took plenty of notes. Overall, the interviews were done between December 2022 and March 2023.
 

\subsection{Main Findings}
\label{ssec:interviews_results}
\noindent
After carrying out the interviews, we qualitatively analyzed all the notes taken (we cannot share such notes due to NDA). We organize our main findings in three areas (summarized in Table~\ref{table:interview_results}): challenges of data labeling, possible improvements (in the short-term), and avenues for future work. Let us present each of these at a high level.

\vspace{-2mm}

\begin{table}[!htbp]
    \centering
        \caption{Interview results.}
    \label{table:interview_results}
    \vspace{-3mm}
    \small
    \noindent\begin{tabular}{@{}|l|l|l|}
        \hline
        \textbf{Challenges} & \textbf{\begin{tabular}[c]{@{}l@{}}Suggested Improvements\end{tabular}} & \textbf{\begin{tabular}[c]{@{}l@{}}Future Work\end{tabular}} \\
        \hline
        \rowcolor[HTML]{DAE8FC} 
        Sensitive Data & Iterative Labeling & \begin{tabular}[c]{@{}l@{}}Self-explainable\\  ML Models\end{tabular}  \\
        Time Expenditure & Active Learning & Early labeling\\
        \rowcolor[HTML]{DAE8FC} 
        Financial Costs & \begin{tabular}[c]{@{}l@{}}Integration of data labeling\\ into Company Routines\end{tabular} &\\
        Lack of Ground Truth &  &\\
        \rowcolor[HTML]{DAE8FC} 
        Continuous Process &  &\\
        Manual Task &  &  \\
        \hline
    \end{tabular}\\

\end{table} 

\textbf{Challenges of Data Labeling}. Our SMEs admitted to facing many challenges during their daily routines w.r.t. data labeling. Among these, we mention the following six.

\begin{itemize}[leftmargin=0.4cm]
    \item\textit{Sensitive data}: Labeling sensitive data poses challenges due to strict privacy regulations. Systems that ensure no direct human interaction with the data are needed to maintain confidentiality.

    \item\textit{Time expenditure}: The time spent on data labeling varies depending on the data type and system dynamics. The complexity is increased by system changes within a short period. Estimating the exact percentage of time spent on labeling is difficult but it consumes a significant amount of time, especially in supervised methods for threat modeling or identifying malicious patterns.
    \item\textit{Costs of data labeling}: Data labeling constitutes a significant portion of the overall costs associated with developing an ML model. However, discussing specific cost numbers is challenging and SMEs have limited knowledge about the costs due to its ongoing nature and budget allocations. 
    \item\textit{Lack of ground truth}: especially for some cyberthreats (e.g., APT~\cite{lemay2018survey}), it is hard even for a SME to provide a reliable label (i.e., is an attack taking place or not?). Iterative labeling is necessary due to difficulties in differentiating between different threat types and the discovery of new patterns. 
    \item\textit{Ongoing nature of data labeling}: Data labeling is an ongoing process due to software updates and changes in the environment or threat landscape. Labeled datasets become obsolete and re-labeling is necessary based on factors such as data type and problem dynamics. This issue is often aggravated by the (well-known) likely ``alert fatigue''~\cite{alahmadi202299}.
    \item\textit{Manual labeling}: Human expertise is crucial in the labeling process. Involving domain experts with a deep understanding of the cybersecurity domain is necessary for accurate labeling.
\end{itemize}
Perhaps interestingly, our interviewees never mentioned ``crowdsourcing'' (§\ref{ssec:background}). This may be because their security companies handle sensitive data that cannot be offloaded to third parties.

\textbf{Improvement Possibilities}: According to our interviewees, there are ways to improve the current process of labeling in the cybersecurity domain. Three, in particular, were identified,
\begin{itemize}[leftmargin=0.4cm]
    \item \textit{Iterative labeling}: Iterative labeling is beneficial for cyberthreat detection. This approach allows continuous adaptation, although it requires reevaluation when new discoveries arise.
    \item \textit{Active learning}: Most companies were not familiar with the concept of active learning (AL). However, those who were aware\footnote{Interestingly, SOC analysts did not know the term, but were unconsciously using AL since alerts in a SOC can be labeled and are also provided with a `confidence' score.} of AL recognize its potential to enhance the efficiency and effectiveness of the data labeling process. 
    \item \textit{Integration into company routines}: Companies strive to seamlessly integrate data labeling into their routines to ensure accurate labeling. However, widespread implementation and standardized processes are still lacking.
\end{itemize}
We anticipate that the findings above inspired us to perform our experimental campaign focused on active learning.

\textbf{Looking ahead.} 
Our interviewees made two intriguing remarks that may revolutionize the process of data labeling in cybersecurity.
\begin{itemize}[leftmargin=0.4cm]
    \item Data labeling in cybersecurity will be influenced by developments in \textit{AI explainability}. Currently, many ML models operate as black boxes, lacking transparency in their decision-making processes. This hinders trust in the outputs of these models and the ability to justify security decisions to stakeholders. The demand for more explainable ML models in cybersecurity is growing, potentially reducing the reliance on human experts for data labeling. Models that can provide credible explanations for their decisions may eliminate the need for human verification. In contrast, more advanced models could offer additional context to assist human experts, reducing the time required for labeling.
    
    \item \textit{Commencing data labeling early} is crucial for cost efficiency and improved learning. Even if better labeling methods emerge in the future, starting with pre-labeled data prevents starting from scratch. Early data labeling expedites the learning cycle and enhances the quality of labeling, leading to long-term cost savings. Companies should ensure their systems allow easy validation or dismissal of data with a single click. However, they must also guard against dismissing results without thorough scrutiny. Proper data labeling is essential to avoid future complications. Effectively mastering data labeling is an ongoing process involving labeling, learning from errors, and repeating the cycle. Therefore, it is advantageous for companies to initiate data labeling as early as possible.
\end{itemize}

\vspace{1mm}

\textbox{
\textbf{Disclaimer:} The statements above stem from our own re-elaboration of the (sparse and unstructured) answers we received during our interviews, and they reflect the opinion of SMEs. 
}

\subsection{Interpretation and Takeaways} 
\label{ssec:interviews_interpretation}

\noindent
We now attempt to interpret the responses received by our interviewees, aiming to derive some actionable takeaways. 

First, \textbf{there is a huge gap between scientific research and industrial practice}. This is evidenced by the following:
\begin{itemize}[leftmargin=0.4cm]
    \item Most of our interviewees did not know the term ``active learning'', despite being common in related literature~\cite{apruzzese2022sok, ren2021survey}.
    \item Despite labeling playing a crucial role in ML development (which had been known for decades~\cite{sommer2010outside,mjolsness2001machine}), companies do not have established workflows for doing so in practice. 
    \item Reaching out to practitioners was hard, since only 5 out of 40 companies accepted to participate in our interviews (a problem encountered also by other studies~\cite{meyer2022smartgrid}).
\end{itemize}
Given the above, our paper is a step in the right direction.

Second, \textbf{the costs of labeling are unknown to both researchers and practitioners alike}. Whenever we asked an interviewee to provide an estimate of such costs (either in terms of allocated resources, or time spent labeling) we have never received a clear answer. What we find intriguing is that our interviewees belonged to large and well-known security companies---and we were expecting that the labeling workflow was at least somewhat structured; in contrast, the reality is that (at least according to our interviewees) these procedures are done manually and occasionally. Hence, we advocate for companies to take the problem of labeling more seriously (even their employees are implicitly requesting it!).

Third, \textbf{labeling is not easy even for SMEs}. This finding is crucial, especially in the cybersecurity context, since it suggests that -- in reality -- the data used to train operational ML models may present abundant errors. As such, from the perspective of a researcher, assuming that a given `benchmark' dataset contains labels that are 100\% accurate (which is the de-facto standard\footnote{This can explain why the near-perfect accuracies of ML models in research environments still trigger skepticism in security practitioners~\cite{apruzzese2023sok}.} in ML papers) may be overly optimistic. We argue that to represent a more realistic scenario, it is necessary to synthetically create some `polluted' data-points which can simulate human labeling errors.

We conclude this section by reporting some insightful remarks that, despite being orthogonal to data labeling, provide further evidence of the `disconnection' between research and practice in the context of ML security (and which complement~\cite{alahmadi202299}).

\begin{cooltextbox}
\textsc{\textbf{Reports from SOC analysts.}}
According to some SMEs who are familiar with research, there is a discrepancy between academic assertions and the practical use of ML, particularly of ``Deep Learning'', in cybersecurity. Despite recent studies claiming extensive application of Deep Learning (e.g.,~\cite{mahdavifar2019application}), the insights from SMEs suggest otherwise. Indeed, according to SMEs, a significant portion of cyberthreat detectors relies on rule-based methods, with ML being used only for complex scenarios.\footnote{Allegedly, ML is only applied to approximately 20\% of the incoming samples, while rules govern decision-making for around 70\%.} Furthermore, when anomalies or changes occur in the network environment, SMEs emphasize the need to revert to simpler ML models and start afresh. This highlights the importance of efficient labeling and raises concerns about the efficacy of deep learning methods, which require larger (labeled) datasets. 
\end{cooltextbox}

\section{User Study}
\label{sec:user}

\noindent
Drawing on the insights gleaned from the interviews (§\ref{sec:interviews}), we carry out a user study to further elucidate the key issues tackled by this paper. We begin by describing the adopted methodology (§\ref{ssec:user_method}), and then present the results (§\ref{ssec:user_results}). 

\subsection{Method}
\label{ssec:user_method}

\noindent
Contrary to the interviews, the user study entails a semi-structured questionnaire~\cite{adeoye2021research}, meant to be answered asynchronously by a different set of participants. 

\textbf{Questionnaire.}
We designed a questionnaire having 15 questions inquiring about the role of labeling (within the participant's company) and about predictions on future developments of ML. Each question is accompanied by a set of 3 to 5 potential answers or a designated space for respondents to provide a custom response. The last three (out of 15) questions are formulated as open prompts. We provide our questionnaire in our repository~\cite{ourRepo}. To encourage participation and ensure a reasonable completion time, the user study is designed to be completed within five minutes. Participants were always given the possibility of not answering some questions.

To ensure consistency and comparability of the results, the questionnaire begins by clarifying the meaning of ``manual labeling'', and by defining the term ``project''\footnote{We defined a project as ``the development of an ML model that yields appreciable detection performance after its deployment''.} (this term occurs frequently in the questionnaire). The first questions (Q) focus on uncovering the role of labeling in the participants' daily routines; for instance, Q3 asks ``what percentage of the whole project is dedicated to labeling?'' with possible answers being ``less than 10\%/between 10\% and 20\%/between 30\% and 50\%/more than 50\%''. We also inquire about a participant's opinion on ``active learning''. The last questions ask about expectations on the future role of supervised ML and explainability in the cybersecurity domain (both of which are linked to data labeling). We hosted the questionnaire on an unpublished website, and we never asked for participants' sensitive or personally identifiable information (the questionnaire is anonymous).

\textbf{Participants.}
We set ourselves the goal of having an increased number of participants for the user study (w.r.t. the expert interviews). Hence, between March and June 2023, we reached out again to dozens of SMEs, each representing a unique security company, asking if they were willing to partake in a short survey about their labeling practices. We restricted all communication to emails: as soon as an SME agreed to participate in our research, we sent them a link to the questionnaire. Since the procedure was asynchronous, we do not know the identity of our respondents (who may have delegated colleagues with better expertise). Nonetheless, by the end of June 2023, we received 13 responses to our user study---all belonging to SMEs representing different security companies.\footnote{We believe that the higher response rate w.r.t. the interviews to be due to the survey being less time-consuming ($\sim$2 minutes) than the interview (1+ hour).}

\begin{figure*}[!htbp]
    \centering
    \begin{subfigure}{0.66\columnwidth}
        \centering
        \includegraphics[width=1\columnwidth]{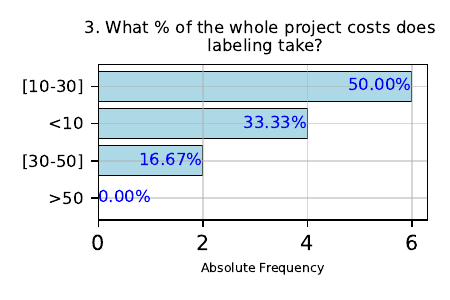}
        \caption{Responses to Q3.}
    \end{subfigure}
    \begin{subfigure}{.66\columnwidth}
        \centering
        \includegraphics[width=1\columnwidth]{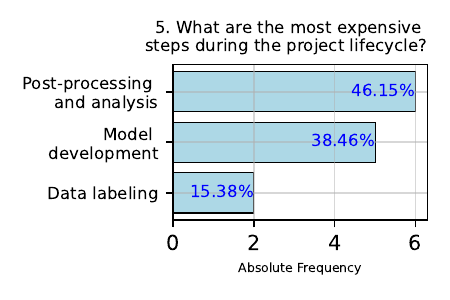}
        \caption{Responses to Q5.}
    \end{subfigure}
    \begin{subfigure}{.66\columnwidth}
        \centering
        \includegraphics[width=1\columnwidth]{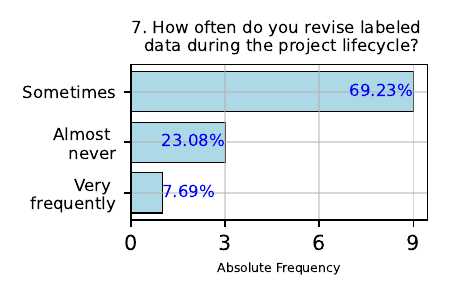}
        \caption{Responses to Q7.}
    \end{subfigure}
    \vspace{-2mm}
    \caption{Some responses of our user study (we provide the full results in our public repository~\cite{ourRepo}).}
    \label{fig:responses}
\end{figure*}

\subsection{Results and Takeaways}
\label{ssec:user_results}
\noindent
With few exceptions, all 13 SMEs responded to every question. We display the responses for three questions (Q3, Q5, Q7) in Fig.~\ref{fig:responses}; the full results are in our repository~\cite{ourRepo}. Due to space limitations, we only discuss the three most relevant findings.

First, Q3 (Fig.~\ref{fig:responses}a) shows varying perspectives on the time dedicated to data labeling, with some SMEs estimating it to be more than 30\% of a project's life-cycle, while others believed it to be between 10 and 30\%, or even less than 10\%. Intriguingly, Q1 reveals that the projects of 54\% of participants take [4--6] months, while those of 31\% take more than 6 months, and the remaining 15\% have projects that last [1--4] months (this supports the observation in~\cite{apruzzese2023sok} that cybersecurity is mostly outsourced). However, Q5 (Fig.~\ref{fig:responses}b) indicates that labeling is (overall) ``less expensive'' than other maintenance operations such as post-processing and analysis. This finding aligns with Q7 (Fig.~\ref{fig:responses}c), which reveals that most practitioners hardly revise previous labels. Hence, we conjecture that \textbf{labeling is done mostly at the beginning} and that -- while important -- after the ML model has been developed, labeling duties are overshadowed by other tasks (we invite the reader to look at Q6 in our repository).

Second, about active learning (in Q10), 4 (31\%) of our respondents have ``never heard of it'', while 4 (31\%) ``are using it'' and 5 (38\%) are ``using something similar''. Further, some remarked that \textbf{AL can lead to overconfidence}, because the expert will only inspect the samples suggested by the ML model, thereby potentially overlooking samples that conceal traces of serious threats. This observation is intriguing and may explain why AL is not yet widespread in security---which is a field in which even a single misclassification can lead to an entire system becoming compromised.

Third, about future prospects, Q15 reveals that our participants have mixed views on the popularity of supervised ML (w.r.t. unsupervised ML): 5 (38\%) participants believe that ``it is unlikely that more supervised ML will be deployed'', whereas the remaining predicted that it would be either ``very popular'' (3, 23\%) or ``more used, but not by much'' (5, 38\%). Regardless, \textbf{the expectation is that supervised ML will remain used in cyberthreat detection}, which urges the development of effective labeling strategies.

\section{Technical Experiments}
\label{sec:experiments}
\noindent
As a last contribution, we now perform proof-of-concept experiments revolving around some of our prior findings. We first examine the impact of various amounts of training data on the performance of a (supervised) ML-based detector (§\ref{ssec:baseline}). Then, we investigate how human labeling errors can affect the quality of an ML model~(§\ref{ssec:mislabeling}). Finally, we assess the benefits of active learning (§\ref{ssec:active}).

\textbf{Dataset.} The cyberthreat detection landscape is large, since it encompasses, e.g., malware, phishing, and network intrusion detection---all of which being domains for which many ML-ready datasets exist~\cite{apruzzese2022sok}. However, recent studies revealed that publicly available datasets for network intrusion detection are flawed~\cite{engelen2021troubleshooting}, whereas many recently proposed malware detectors in research entail deep learning---which our SME regarded with skepticism (see end of §\ref{ssec:interviews_interpretation}). Hence, we focus these experiments on the problem of \textit{phishing website detection}, given that (i)~ many works (e.g.,~\cite{montaruli2023raze,tian2018needle,apruzzese2022mitigating}) showed that ``shallow'' ML algorithms outperform those based on deep neural networks; and that, for phishing detection, (ii)~we are more confident that the labels are correct~\cite{apruzzese2022sok}. Among the many datasets containing phishing and benign websites (e.g.,~\cite{apruzzese2022mitigating, montaruli2023raze}), we chose the well-known one by Chiew et al.~\cite{chiew2019new}. It contains \smamath{10\,000} samples (webpages), equally split between benign (taken from Alexa top) and phishing (taken from OpenPhish and PhishTank).

\vspace{1mm}

\textbox{\textbf{Disclaimer:} The goal of our evaluation is to guide future research by suggesting some ``viewpoints'' often neglected in related literature (but relevant in practice). We do not claim technical novelty, but our results can serve as a benchmark (we share our code~\cite{ourRepo}, which also includes the hyperparameters and low-level details).}

\subsection{Training Size Impact (Baseline)}
\label{ssec:baseline}
\noindent
As a starting point, we study the performance of an ML-detector as a function of the amount of labeled data used during its training phase. While similar studies have been carried out in the past (even for phishing 
 website detection---e.g.,~\cite{apruzzese2022sok, li2017phishbox, yang2017multi}), we are not aware of works that performed such an evaluation on our chosen dataset. Furthermore, related papers on phishing detection perform their experiments on ``private'' data (such as~\cite{li2017phishbox,yang2017multi}). Hence, our testbed represents a valuable benchmark for future work.

\textbf{Setup.}
We embrace the recommendations of~\cite{apruzzese2022sok}. First, we take our dataset~\cite{chiew2019new} (having 5k benign/phishing samples), \smabb{D}, and extract a test partition, \smabb{E}, containing \smamath{20\%} of \smabb{D} (i.e., 1k benign/malicious samples). The remaining \smamath{80\%} samples of \smabb{D} are then treated as data usable for training, \smabb{T}. For developing our ML detector, we rely on the random forest (RF) algorithm---which has been shown to consistently outperform other types of classification algorithms for phishing website detection (e.g.,~\cite{tian2018needle, montaruli2023raze, li2017phishbox, chiew2019new}). Since in this experiment we want to measure the impact of different amounts of labeled data, we train ML models by randomly sampling from \smabb{T} at \smamath{1\%} increments, spanning from 1\% (i.e., \smamath{80} labeled samples) to 100\% of \smabb{T} (i.e., \smamath{8\,000} samples); for consistency, we ensure that every subset is balanced. Hence, for every considered subset of \smabb{T}, we train an RF classifier and assess its performance on \smabb{E}. Finally, to provide a statistically significant benchmark, we repeat the sampling \smamath{30} times (i.e., we develop \smamath{3\,000} ML models: 30 trials\smamath{\times}100 subsets of \smabb{T}): according to~\cite{apruzzese2022sok}, simulations of scarce amounts of labeled data (drawn from a large labeled dataset) may present sampling bias which must be accounted for by repeating the draw many times (which is not done, e.g., in~\cite{yang2017multi}). Inspired by this observation, we will \textit{compare the results of a `single' trial with the results (averaged) of \smamath{30} trials}. We measure the performance of every assessment via common evaluation metrics, i.e., accuracy, recall, precision, and F1-score; for simplicity, we only consider accuracy in this section (we report the other metrics in our repository~\cite{ourRepo}).

\begin{figure}[!htbp]
    \centering
    \includegraphics[width=\columnwidth]{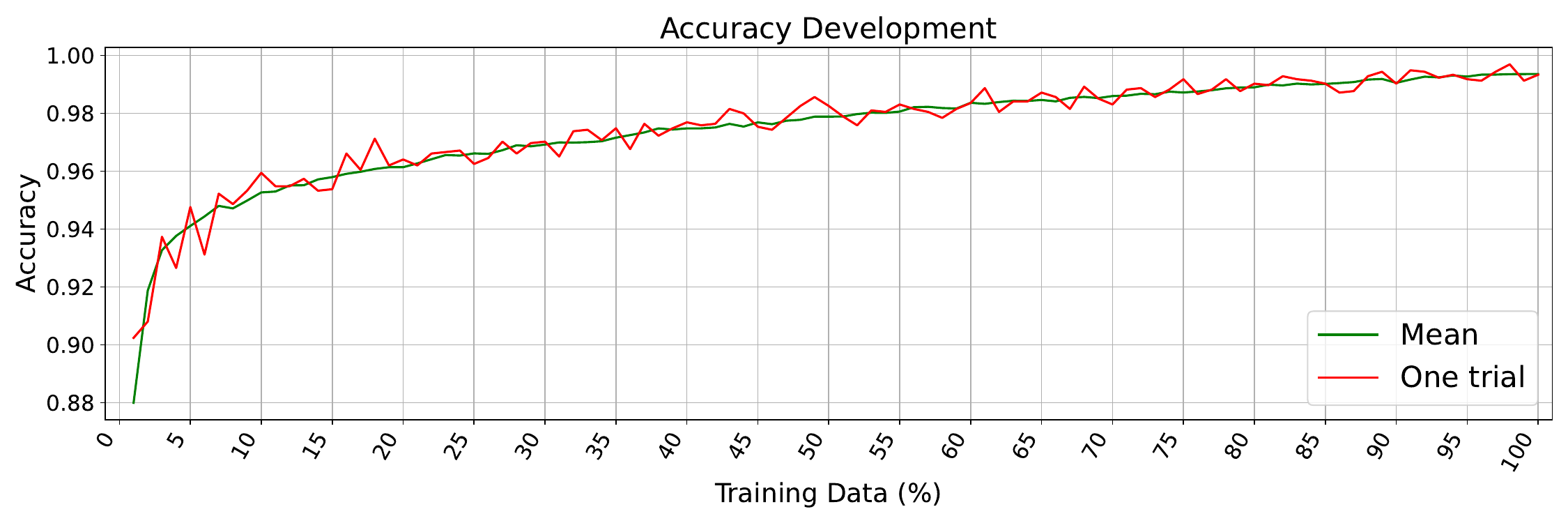}
    \vspace{-5mm}
    \caption{Performance as a function of the training size. \textmd{\small We further compare the average performance (over 30 trials) w.r.t. a single run.}}
    \label{fig:baseline}
\end{figure}

\textbf{Results.}  We visualize the results of this experiment in Fig.~\ref{fig:baseline}, showing the performance (y-axis) as a function of the size of the training set (x-axis). The green line refers to the average accuracy over the \smamath{30} trials, whereas the red line refers to the accuracy of a single (randomly chosen) trial. By observing the green line, we can see that the accuracy (we recall that \smabb{E} is a balanced dataset) is already above \smamath{88\%} with only \smamath{1\%} of \smabb{T}, and it reaches \smamath{95\%} with \smamath{12\%} of \smabb{T}. This is an intriguing result (which partially echoes those in~\cite{li2017phishbox,apruzzese2022sok}) since it shows that (at least on this dataset) it is not necessary to resort on a large labeled dataset to develop proficient ML detectors. Nonetheless, by observing the red line, we see an inconsistent trend (w.r.t. the stable one of the green line): e.g., for the red line, the accuracy for \smamath{18\%} of \smabb{T} is \smamath{97\%} whereas the green line reaches such a value only for \smamath{35\%} of \smabb{T}. This underscores the importance of carrying out multiple trials, since a single `lucky' draw may yield to overly-optimistic performance. Finally, we also note that the average accuracy with 100\% of \smabb{T} (i.e., \smamath{80\%} of the original \smabb{D}) is \smamath{\sim}\smamath{99\%}, a result that aligns with the one by the creators of \smabb{D}~\cite{chiew2019new} (which confirms the quality of our implemented ML detector).

\vspace{-3mm}
\vspace{-3mm}

\begin{cooltextbox}
\textsc{\textbf{Takeaways.}}\footnote{These only apply to our dataset and ML algorithm, and we do not generalize.} First, large labeled datasets are not always necessary to yield appreciable performance---hence we endorse researchers to experiment with smaller amounts of labeled data. Second, when randomly sampling from small datasets, the performance between multiple and single trials is remarkably different---hence we encourage researchers to repeat their assessments.
\end{cooltextbox}

\begin{figure*}[!htbp]
    \centering
    \begin{subfigure}{0.99\columnwidth}
        \centering
        \includegraphics[width=1\columnwidth]{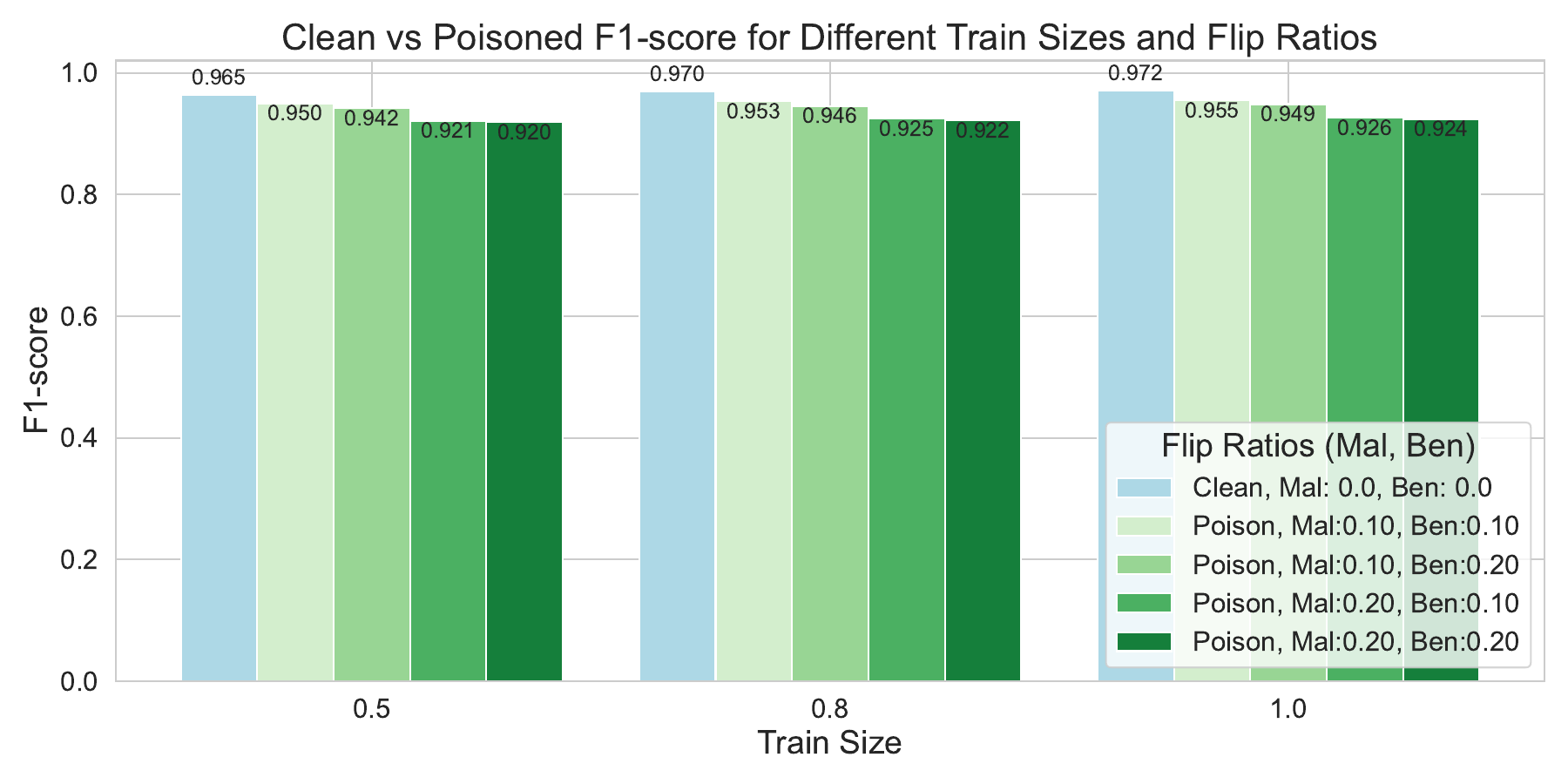}
        \caption{Effect on F1-score (averaged over 30 trials).}
    \end{subfigure}
    \begin{subfigure}{.99\columnwidth}
        \centering
        \includegraphics[width=1\columnwidth]{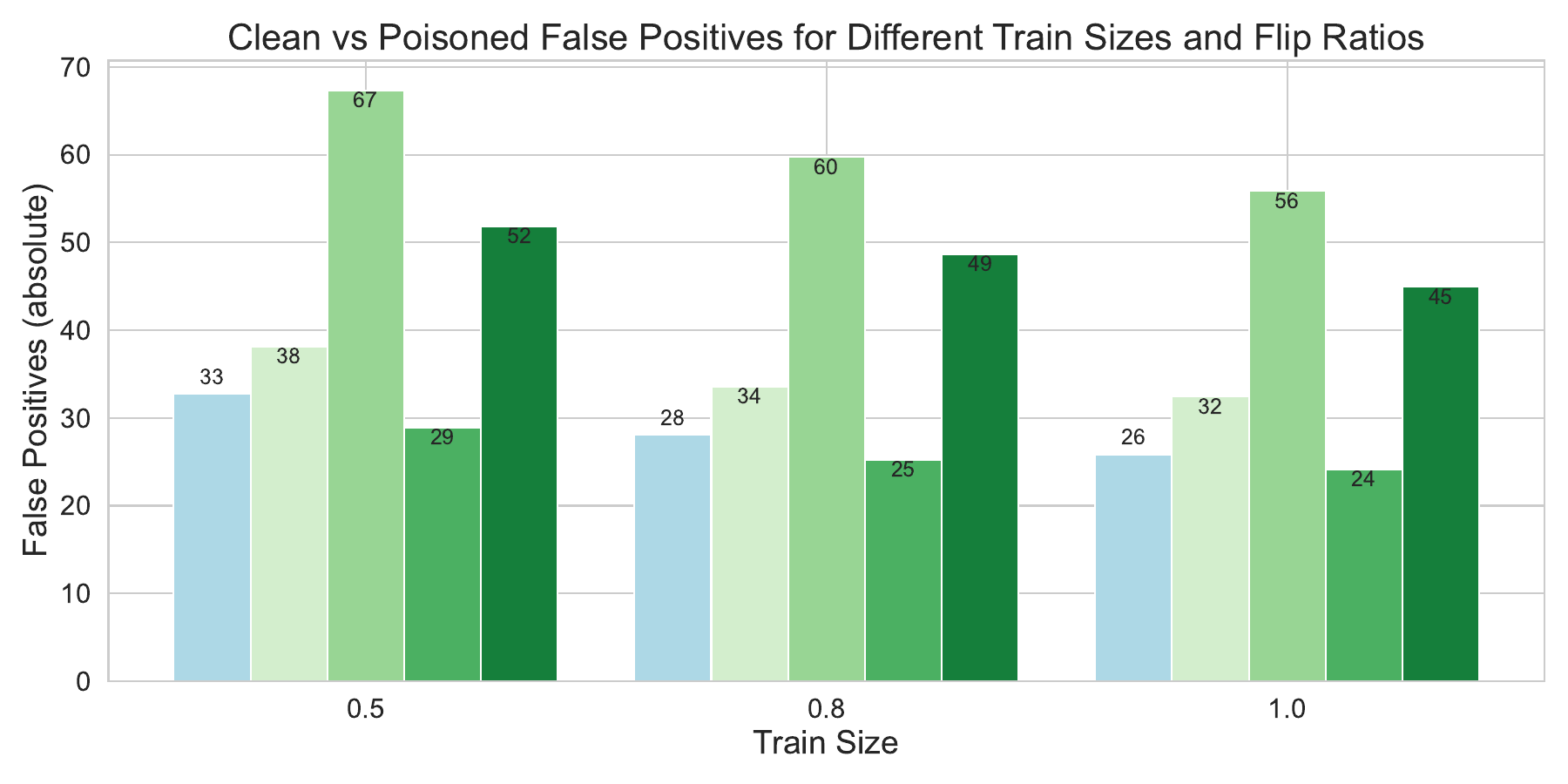}
        \caption{Effect on False Positives (absolute---averaged over 30 trials).}
    \end{subfigure}
    \vspace{-2mm}
    \caption{Impact of mislabeling. \textmd{\small We simulate human error by flipping the label of some subsets of the training data to see how much the performance changes.}
    }
    \label{fig:mislabeling}
\end{figure*}

\subsection{Human Error Impact}
\label{ssec:mislabeling}
\noindent 
Next, inspired by some of our findings (revealing that labels may be revised---see §\ref{ssec:user_results}), we seek to evaluate what happens if some of the labels in a given training dataset are incorrect.

\textbf{Setup.} We follow a similar procedure as in the previous experiment (see §\ref{ssec:baseline}). The first difference is that we consider only three (instead of a hundred) subsets of \smabb{T}: \smamath{50\%}, \smamath{80\%}, \smamath{100\%} (i.e., \smamath{40\%}, \smamath{64\%}, \smamath{80\%} of the original \smabb{D}). The second difference entails the method we use to simulate the erroneous labels. Specifically, for any given subset of \smabb{T} used for training, we `flip' the class of some of its labels---thereby inducing some form of ``self-poisoning''~\cite{kan2021investigating}. We consider five \textit{flip ratios} (listed in Table~\ref{table:flip_ratio}), each denoting the percentage of samples of a given class (benign or malicious) whose label is flipped (i.e., a `benign' sample is assigned to a `malicious' label, and vice-versa). We repeat all experiments 30 times, averaging the performance---always measured on the same \smabb{E} (\smamath{20\%} of \smabb{D}) for consistency.

\vspace{-2mm}
\begin{table}[!htbp]
    \centering
    \caption{Flip Ratios---\textmd{\small 
     We simulate poisoning by flipping the label (`benign' becomes `malicious', and vice-versa) of a subset of the training data.}}
    \vspace{-3mm} 
    \resizebox{0.5\columnwidth}{!}{%
    \begin{tabular}{cccc}
    \toprule
     & \textbf{Malicious} & \textbf{Benign} & \textbf{Poisoning} \\
    \midrule
    \textbf{Clean} & 0.0 & 0.0 & 0\% \\
    \midrule
    \multirow{4}{*}{\textbf{Poison}}
     & 0.10 & 0.10 & 20\% \\
     & 0.10 & 0.20 & 30\% \\
     & 0.20 & 0.10 & 30\% \\
     & 0.20 & 0.20 & 40\% \\
    \bottomrule
    \label{table:flip_ratio}
    \end{tabular}
    }
\end{table}
\vspace{-3mm}

\textbf{Results.} We visualize the results in Fig.~\ref{fig:mislabeling}, showing the performance (y-axis) for each training subset of \smabb{T} (group of bars), and for each flip ratio (individual bars); specifically, the light-blue bar is `clean' (which we use as baseline) and the green ones entail `poisoning'. Fig.~\ref{fig:mislabeling}(a) focuses on F1-score, whereas Fig.~\ref{fig:mislabeling}(b) on the (absolute) false positives.
By observing Fig.~\ref{fig:mislabeling}(a), an intriguing result is that, despite the substantial difference in training data size, the performance barely changes (i.e., each bar of the same color has very similar F1-scores---confirmed by a two-sample statistical t-test). Furthermore, another surprising observation is that the effect on the F1-score is mild: even when \smamath{40\%} of the labeled data is poisoned (dark-green bar), the drop w.r.t. the baseline (light-blue bar) is minor (from \smamath{0.97} to \smamath{0.92}). However, a two-sample statistical t-test confirms that the drop is statistically significant (\smamath{p}-value \smamath{\approx 0}). 

In contrast, by focusing on Fig.~\ref{fig:mislabeling}(b), we see an almost contradictory result: the highest number of false positives is always achieved by the middle bar in each group---which is \textit{not the one with the largest percentage of poisoning} (which is the rightmost bar in each group). Given that in phishing detection false positives tend to be very annoying to end-users, this reveals that practitioners should be extra-cautious when assigning `malicious' labels (indeed, the middle bar has \smamath{20\%} of benign samples being turned into malicious ones, and \smamath{10\%} of malicious samples being turned into benign ones).

\begin{cooltextbox}
\textsc{\textbf{Takeaways.}}\footnote{These only apply to our dataset and ML algorithm, and we do not generalize.} For certain amounts of training data size, mislabeled data (poisoning) leads to negligible performance differences. However, while the F1-score appears to be a `robust' metric to poisoning, the amount of false positives is affected by a specific type of poisoned classes. Hence, we endorse researchers to pay more attention to the poisoned class. For mitigations, see, e.g.,~\cite{chan2022fault}.
\end{cooltextbox}

\subsection{Active Learning Gain}
\label{ssec:active}
\noindent
Lastly, we turn the attention to active learning (AL) due to the `mixed' viewpoint expressed by our SME on this technique (see §\ref{ssec:interviews_results} and §\ref{ssec:user_results}). We recall that extensive background on AL is in Appendix~\ref{app:active}, which discusses the specific method of \textit{uncertainty sampling}---which we will use in our experiments due to its simplicity and demonstrated effectiveness (e.g.,~\cite{li2017phishbox,apruzzese2022sok,pendlebury2019tesseract}).

\textbf{Setup.} 
We adopt a similar setup as in §\ref{ssec:baseline}. The implementation of AL is similar\footnote{For this experiment we changed the RF algorithm by setting the `bootstrap' option to False (setting it to True, which we did in §\ref{ssec:baseline}, yielded spurious artifacts here).} to the one in~\cite{apruzzese2022sok}. However, the difference lies in our assessment methodology. In particular, we seek to pinpoint the performance gain by assuming a fixed labeling budget but spread over many iterations. To give an idea, assume that an annotator has a labeling budget of 50 samples. In~\cite{apruzzese2022sok} (or also in~\cite{li2017phishbox}) the authors compared the gain using, e.g., 50 randomly chosen labeled samples w.r.t. 50 actively suggested labeled samples. In contrast, we want to compare what happens if the 50 actively suggested labeled samples derive from the annotator labeling such samples ``all together'' w.r.t. doing so by splitting the labeling task in ``mini-batches''. I.e., labeling a subset of these 50 (e.g., 25), and then using such batch to update the ML model which is then used to provide ``updated suggestions'', which will finally be provided to the annotator for another (or more) round of labeling. To do this, we `fix' the labeling budget to the entire \smabb{T}, and then consider different 5 different amounts of labeling iterations, i.e., [2,4,16,32,64]; the first iteration is always randomly chosen (this is a realistic assumption---§\ref{ssec:interviews_results} and~\cite{apruzzese2022sok}). After each iteration, we measure the performance of the resulting updated ML model (always on the same \smabb{E}, for consistency). As usual, we repeat these experiments 100 times for statistical robustness.\footnote{\textbf{For example}, 4 iterations means that \scbb{T} (having 8k samples) is first randomly sampled (by drawing 2k samples) used to train an initial ML model \scmath{M}; this \scmath{M} is then used to suggest 2k samples to an annotator and, after being correctly labeled, will be used to update \scmath{M} (which is now trained over 4k samples). The updated \scmath{M} will then further suggest 2k samples to label, and then be updated again (now with 6k samples). The procedure will be repeated  one last time---thereby exhausting the labeling budget of 8k samples (i.e., the full \scbb{T}). After every update of \scmath{M}, we assess its performance on \scbb{E}. We repeat this process 10 times, and then select a new \scbb{E} (as recommended by~\cite{apruzzese2022sok}) and start again for 10 more times. We average all results. (\textbf{Runtime} is in Appendix~\ref{app:runtime}.)}

\vspace{-2mm}

\begin{figure}[!htbp]
    \centering
    \includegraphics[width=\columnwidth]{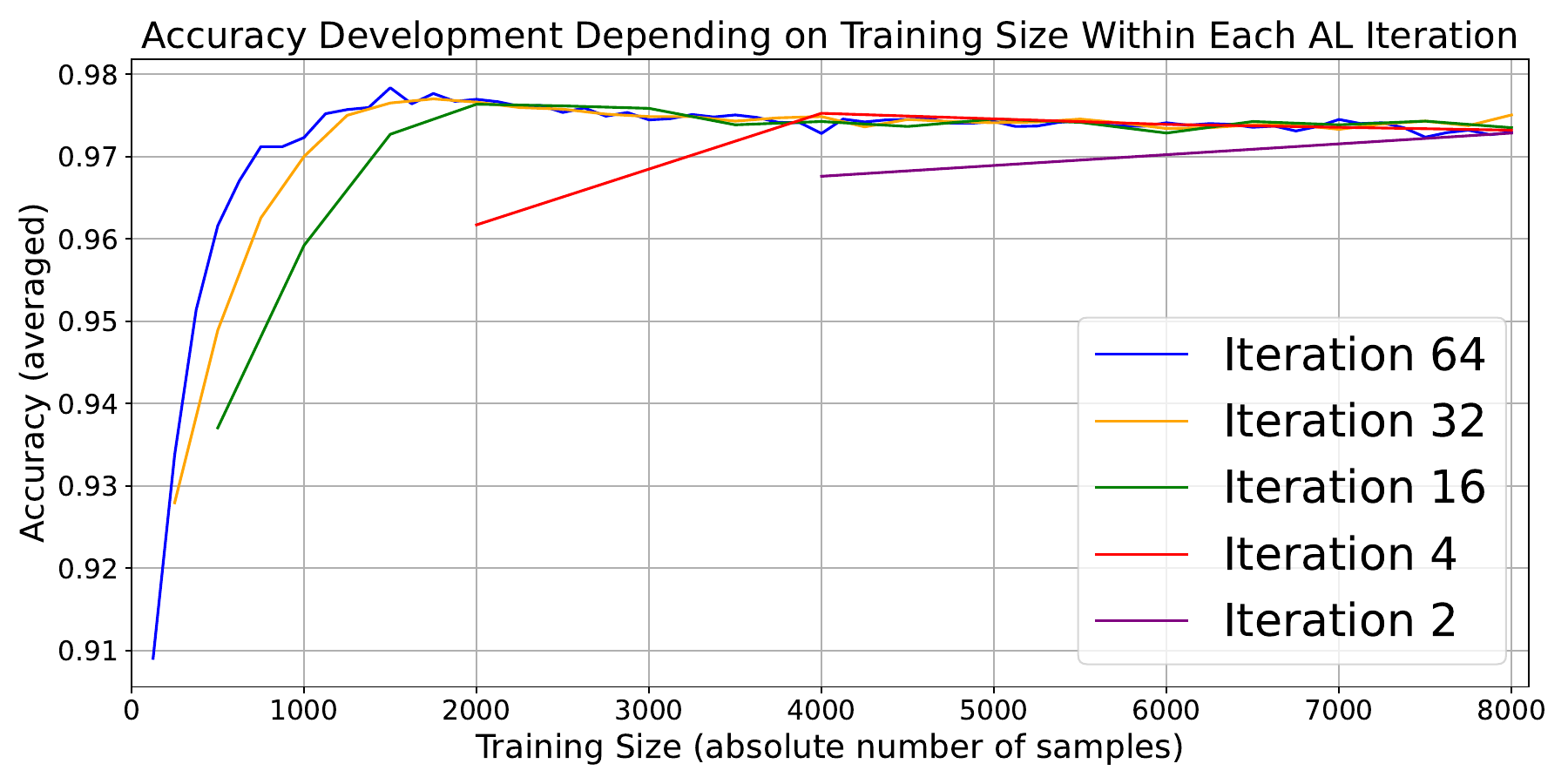}
    \caption{Impact of Active Learning (avg 100 trials). \textmd{\small We compare the gains of labeling the suggested samples ``all together'' w.r.t. doing so over many iterations---each done by updating the model and suggesting new samples.}
    }
    \label{fig:active}
\end{figure}

\vspace{-2mm}

\textbf{Results.} We report the results in Fig.~\ref{fig:active}, which shows the Accuracy (y-axis) as a function of the amount of the training size (x-axis); lines indicate the specific amount of iterations. As usual, we report further visualizations entailing other performance metrics (F1-score, Recall, Precision) and training sizes (\smamath{50\%} and \smamath{80\%} of \smabb{T}) in our repository~\cite{ourRepo}. By observing Fig.~\ref{fig:active}, we see that the first iteration (which reports the performance after randomly choosing samples) is always much worse than the following ones---which is expected. 
However, we make an intriguing observation: there is a stark difference between the `gain' of multiple labeling tasks, w.r.t. fewer ones. To appreciate this, we consider two use cases. \textbf{(1) Scarce Budgets}, i.e., when an annotator can label at most 1.5k samples. Let us focus our attention on the blue, yellow, and green lines. For the green line, after one round of active learning (500 random and 500 suggested labels over a single round), the Accuracy is \smamath{96\%}; for the yellow line, after four rounds (250 random, and 750 suggested labels over three rounds) the Accuracy is \smamath{97\%}; for the blue line, after eight rounds (125 random, and 875 suggested labels) the Accuracy is \smamath{97.2\%}. This may indicate that splitting the labeling task into multiple batches may be advantageous. However, this is not true for \textbf{(2) Abundant Budgets}, i.e., when an annotator can afford more than 4k labels. Let us compare the blue with the orange line. For the orange line, after one round of active learning, (2k random and 2k suggested labels), the Accuracy is \smamath{97.5\%}. For the blue line, after 32 rounds (125 random and 3875 suggested), the Accuracy is \smamath{97.3\%}. Indeed, after a certain point, the performance saturates and there is no gain in splitting the labeling into multiple batches.

\begin{cooltextbox}
\textsc{\textbf{Takeaways.}}\footnote{These only apply to our dataset and ML algorithm, and we do not generalize.} Applying active learning (through uncertainty sampling) by splitting the labeling task into multiple batches is advantageous in the initial development phases, but yields no returns after some saturation points. We endorse researchers to identify these plateaus in other domains and datasets.
\end{cooltextbox}
\section{Discussion}
\label{sec:discussion}

\textbf{Comparison with Prior Work.}
Our paper shares similarities with prior (peer-reviewed) works that touch the problem labeling in practice. Here, we discuss two of these. 
\textbf{(1)} Fredriksson et al.~\cite{fredriksson2020data} carry out interviews (in 2019) with five SMEs belonging to two companies (ours belong to five companies) in a single country (ours belong to five countries). However, the role of cybersecurity is unclear: the two companies to which their SMEs belong to are only reported to be ``telecommunication providers'' and ``a company specialized in labeling''. Intriguingly, one participant from the latter company reported that ``labeling takes 200 times less with active learning'', which is in stark contrast to what most of our SMEs reported (both in our interviews and in the user study). 
\textbf{(2)}~Koh et al.~\cite{koh2024voices} conduct semi-structured interviews with 21 ML practitioners (4 of whom work in cybersecurity) on the EU AI Act. While their results do not allow to identify the specific responses of the security practitioners, they found that \smamath{20\%} of their interviewees perform a ``thorough labeling process'', potentially suggesting that some industries do have systematic labeling approaches in place.

\textbf{Implications for AI Security.} Our study reveals that the process of labeling in cybersecurity is still at an early stage and that the many methods proposed in research to deal with this problem are far from being a panacea. This underscores a problem: labeled data is necessary for applications of ML in cyberthreat detection~\cite{apruzzese2022sok}, but the immaturity of the currently adopted labeling practices leads to ML-driven security systems with tradeoffs. E.g., incorrectly labeled data leads to self-poisoning~\cite{kan2021investigating} which degrades performance (as we also showed in §\ref{ssec:mislabeling}). Moreover, the lack of a structured approach to labeling security events also hinders updating the ML model with new (correctly) labeled data, thereby exposing it to evasion attacks~\cite{apruzzese2023role}. Ultimately, this paper is a call to action: by building a bridge between research and practice, novel solutions---at both the technical and organizational levels ---could be developed that improve the reliability (in terms of detection performance and generic robustness) of security systems empowered by ML.

\textbf{Limitations.} We conducted open interviews with five SMEs and carried out a semi-structured user study with 13 SMEs---all operating in the field of cybersecurity and ML, and belonging to different companies. The recruitment of experts with experience in both areas proved challenging due to their scarcity, which is common in related studies~\cite{koh2024voices,alahmadi202299} whose population hardly goes above 20. As such, we do not aim to generalize our findings (but we are not aware of studies focused on security that do so), and our small sample size prevents one from deriving statistically-rooted conclusions. Our experiments are a proof-of-concept and there exist many ways to carry out our evaluation. For instance, we consider only one ML algorithm (despite being the best for the chosen task~\cite{chiew2019new}) and one AL strategy (which is known to work well~\cite{apruzzese2022sok}) and experiment on only one dataset (which is popular~\cite{basit2021comprehensive}) of a subdomain of cyberthreat detection---but this is due to recent works showing the unreliability of prior datasets in other domains~\cite{irolla2018duplication,engelen2021troubleshooting}. We advocate future work to replicate our experiments on different datasets (and we release our tools to facilitate this~\cite{ourRepo})

\section{Conclusions and Recommendations}
\label{sec:conclusions}
\noindent
We investigated the problem of data labeling from the perspective of operational ML security. We interviewed and carried out a user study with security professionals with experience in ML development. Our findings elucidate the hurdles and issues that practitioners face in their daily routines when managing ML-driven security systems. We then carried out technical experiments aimed at showcasing some pragmatic aspects of data labeling which are seldom considered in research on cyberthreat detection. 

To improve the security and robustness of their ML systems while reducing data labeling costs, companies can take the following steps: \textbf{(1)}~\textit{Optimize Label Quality}: Ensure high-quality data labeling from the beginning to avoid costly revisions of previously labeled data, \textbf{(2)}~\textit{Implement Active Learning (AL):} AL can reduce time and cost by achieving high performance with fewer iterations, reducing the amount of data to be labeled, \textbf{(3)}~\textit{Set Stop Criteria for AL Cycles:} After a certain number of iterations, a performance plateau may be reached where further labeling efforts may not be cost-efficient. Setting stop criteria can reduce time and cost, and \textbf{(4)}~\textit{Integrate Data Labeling into Workflows:} Incorporating data labeling into ongoing work processes enhances efficiency and reduces labeling time.

\textsc{\textbf{Acknowledgments.}} We thank the Hilti Corporation for funding, and the practitioners we interviewed for their contributions, availability, and valuable feedback.

\appendix

\bibliographystyle{ACM-Reference-Format}

{


\begin{thebibliography}{64}


\ifx \showCODEN    \undefined \def \showCODEN     #1{\unskip}     \fi
\ifx \showDOI      \undefined \def \showDOI       #1{#1}\fi
\ifx \showISBNx    \undefined \def \showISBNx     #1{\unskip}     \fi
\ifx \showISBNxiii \undefined \def \showISBNxiii  #1{\unskip}     \fi
\ifx \showISSN     \undefined \def \showISSN      #1{\unskip}     \fi
\ifx \showLCCN     \undefined \def \showLCCN      #1{\unskip}     \fi
\ifx \shownote     \undefined \def \shownote      #1{#1}          \fi
\ifx \showarticletitle \undefined \def \showarticletitle #1{#1}   \fi
\ifx \showURL      \undefined \def \showURL       {\relax}        \fi
\providecommand\bibfield[2]{#2}
\providecommand\bibinfo[2]{#2}
\providecommand\natexlab[1]{#1}
\providecommand\showeprint[2][]{arXiv:#2}

\bibitem[our(2024)]%
        {ourRepo}
 \bibinfo{year}{2024}\natexlab{}.
\newblock \bibinfo{booktitle}{\emph{Our Repository}}.
\newblock
\urldef\tempurl%
\url{https://github.com/hihey54/sac24_labeling}
\showURL{%
\tempurl}


\bibitem[Adeoye-Olatunde and Olenik(2021)]%
        {adeoye2021research}
\bibfield{author}{\bibinfo{person}{Omolola~A Adeoye-Olatunde} {and} \bibinfo{person}{Nicole~L Olenik}.} \bibinfo{year}{2021}\natexlab{}.
\newblock \showarticletitle{Research and scholarly methods: Semi-structured interviews}.
\newblock \bibinfo{journal}{\emph{JACCP}} (\bibinfo{year}{2021}).
\newblock


\bibitem[Aghakhani et~al\mbox{.}(2023)]%
        {aghakhani2023venomave}
\bibfield{author}{\bibinfo{person}{Hojjat Aghakhani}, \bibinfo{person}{Lea Sch{\"o}nherr}, \bibinfo{person}{Thorsten Eisenhofer}, \bibinfo{person}{Dorothea Kolossa}, \bibinfo{person}{Thorsten Holz}, \bibinfo{person}{Christopher Kruegel}, {and} \bibinfo{person}{Giovanni Vigna}.} \bibinfo{year}{2023}\natexlab{}.
\newblock \showarticletitle{VenoMave: Targeted poisoning against speech recognition}. In \bibinfo{booktitle}{\emph{IEEE SaTML}}.
\newblock


\bibitem[Alahmadi et~al\mbox{.}(2022)]%
        {alahmadi202299}
\bibfield{author}{\bibinfo{person}{Bushra~A Alahmadi}, \bibinfo{person}{Louise Axon}, {and} \bibinfo{person}{Ivan Martinovic}.} \bibinfo{year}{2022}\natexlab{}.
\newblock \showarticletitle{99\% False Positives: A Qualitative Study of Soc Analysts' Perspectives on Security Alarms}. In \bibinfo{booktitle}{\emph{USENIX Sec}}.
\newblock


\bibitem[Andresini et~al\mbox{.}(2021a)]%
        {andresini2021gan}
\bibfield{author}{\bibinfo{person}{Giuseppina Andresini}, \bibinfo{person}{Annalisa Appice}, \bibinfo{person}{Luca De~Rose}, {and} \bibinfo{person}{Donato Malerba}.} \bibinfo{year}{2021}\natexlab{a}.
\newblock \showarticletitle{GAN augmentation to deal with imbalance in imaging-based intrusion detection}.
\newblock \bibinfo{journal}{\emph{Future Generation Computer Systems}}  \bibinfo{volume}{123} (\bibinfo{year}{2021}), \bibinfo{pages}{108--127}.
\newblock


\bibitem[Andresini et~al\mbox{.}(2021b)]%
        {andresini2021insomnia}
\bibfield{author}{\bibinfo{person}{Giuseppina Andresini}, \bibinfo{person}{Feargus Pendlebury}, \bibinfo{person}{Fabio Pierazzi}, \bibinfo{person}{Corrado Loglisci}, \bibinfo{person}{Annalisa Appice}, {and} \bibinfo{person}{Lorenzo Cavallaro}.} \bibinfo{year}{2021}\natexlab{b}.
\newblock \showarticletitle{Insomnia: Towards concept-drift robustness in network intrusion detection}. In \bibinfo{booktitle}{\emph{2021 ICACT}}. \bibinfo{pages}{111--122}.
\newblock


\bibitem[Apruzzese et~al\mbox{.}(2023b)]%
        {apruzzese2023role}
\bibfield{author}{\bibinfo{person}{Giovanni Apruzzese}, \bibinfo{person}{Pavel Laskov}, \bibinfo{person}{Edgardo Montes~de Oca}, \bibinfo{person}{Wissam Mallouli}, \bibinfo{person}{Luis Brdalo~Rapa}, \bibinfo{person}{Athanasios~Vasileios Grammatopoulos}, {and} \bibinfo{person}{Fabio Di~Franco}.} \bibinfo{year}{2023}\natexlab{b}.
\newblock \showarticletitle{The role of machine learning in cybersecurity}.
\newblock \bibinfo{journal}{\emph{ACM DTRAP}} (\bibinfo{year}{2023}).
\newblock


\bibitem[Apruzzese et~al\mbox{.}(2023a)]%
        {apruzzese2023sok}
\bibfield{author}{\bibinfo{person}{Giovanni Apruzzese}, \bibinfo{person}{Pavel Laskov}, {and} \bibinfo{person}{Johannes Schneider}.} \bibinfo{year}{2023}\natexlab{a}.
\newblock \showarticletitle{SoK: Pragmatic Assessment of Machine Learning for Network Intrusion Detection}. In \bibinfo{booktitle}{\emph{EuroS\&P}}.
\newblock


\bibitem[Apruzzese et~al\mbox{.}(2022a)]%
        {apruzzese2022sok}
\bibfield{author}{\bibinfo{person}{Giovanni Apruzzese}, \bibinfo{person}{Pavel Laskov}, {and} \bibinfo{person}{Aliya Tastemirova}.} \bibinfo{year}{2022}\natexlab{a}.
\newblock \showarticletitle{SoK: The impact of unlabelled data in cyberthreat detection}. In \bibinfo{booktitle}{\emph{IEEE EuroS\&P}}.
\newblock


\bibitem[Apruzzese et~al\mbox{.}(2022b)]%
        {apruzzese2022cross}
\bibfield{author}{\bibinfo{person}{Giovanni Apruzzese}, \bibinfo{person}{Luca Pajola}, {and} \bibinfo{person}{Mauro Conti}.} \bibinfo{year}{2022}\natexlab{b}.
\newblock \showarticletitle{The cross-evaluation of machine learning-based network intrusion detection systems}.
\newblock \bibinfo{journal}{\emph{TNSM}} (\bibinfo{year}{2022}).
\newblock


\bibitem[Apruzzese and Subrahmanian(2022)]%
        {apruzzese2022mitigating}
\bibfield{author}{\bibinfo{person}{Giovanni Apruzzese} {and} \bibinfo{person}{VS Subrahmanian}.} \bibinfo{year}{2022}\natexlab{}.
\newblock \showarticletitle{Mitigating Adversarial Gray-Box Attacks Against Phishing Detectors}.
\newblock \bibinfo{journal}{\emph{TDSC}} (\bibinfo{year}{2022}).
\newblock


\bibitem[Arp et~al\mbox{.}(2022)]%
        {arp2022and}
\bibfield{author}{\bibinfo{person}{Daniel Arp}, \bibinfo{person}{Erwin Quiring}, \bibinfo{person}{Feargus Pendlebury}, \bibinfo{person}{Alexander Warnecke}, \bibinfo{person}{Fabio Pierazzi}, \bibinfo{person}{Christian Wressnegger}, \bibinfo{person}{Lorenzo Cavallaro}, {and} \bibinfo{person}{Konrad Rieck}.} \bibinfo{year}{2022}\natexlab{}.
\newblock \showarticletitle{Dos and don'ts of machine learning in computer security}. In \bibinfo{booktitle}{\emph{USENIX Security}}.
\newblock


\bibitem[Basit et~al\mbox{.}(2021)]%
        {basit2021comprehensive}
\bibfield{author}{\bibinfo{person}{Abdul Basit}, \bibinfo{person}{Maham Zafar}, \bibinfo{person}{Xuan Liu}, \bibinfo{person}{Abdul~Rehman Javed}, \bibinfo{person}{Zunera Jalil}, {and} \bibinfo{person}{Kashif Kifayat}.} \bibinfo{year}{2021}\natexlab{}.
\newblock \showarticletitle{A comprehensive survey of AI-enabled phishing attacks detection techniques}.
\newblock \bibinfo{journal}{\emph{Telecommunication Systems}} (\bibinfo{year}{2021}).
\newblock


\bibitem[Carlini(2021)]%
        {carlini2021poisoning}
\bibfield{author}{\bibinfo{person}{Nicholas Carlini}.} \bibinfo{year}{2021}\natexlab{}.
\newblock \showarticletitle{Poisoning the unlabeled dataset of $\{$Semi-Supervised$\}$ learning}. In \bibinfo{booktitle}{\emph{30th USENIX Security Symposium (USENIX Security 21)}}. \bibinfo{pages}{1577--1592}.
\newblock


\bibitem[Catillo et~al\mbox{.}(2022)]%
        {catillo2022transferability}
\bibfield{author}{\bibinfo{person}{Marta Catillo}, \bibinfo{person}{Andrea Del~Vecchio}, \bibinfo{person}{Antonio Pecchia}, {and} \bibinfo{person}{Umberto Villano}.} \bibinfo{year}{2022}\natexlab{}.
\newblock \showarticletitle{Transferability of machine learning models learned from public intrusion detection datasets: the cicids2017 case study}.
\newblock \bibinfo{journal}{\emph{Software Quality Journal}} (\bibinfo{year}{2022}).
\newblock


\bibitem[Chan et~al\mbox{.}(2022)]%
        {chan2022fault}
\bibfield{author}{\bibinfo{person}{Abraham Chan}, \bibinfo{person}{Arpan Gujarati}, \bibinfo{person}{Karthik Pattabiraman}, {and} \bibinfo{person}{Sathish Gopalakrishnan}.} \bibinfo{year}{2022}\natexlab{}.
\newblock \showarticletitle{The fault in our data stars: studying mitigation techniques against faulty training data in machine learning applications}. In \bibinfo{booktitle}{\emph{Proc. IEEE DSN}}.
\newblock


\bibitem[Chen et~al\mbox{.}(2020)]%
        {chen2020malware}
\bibfield{author}{\bibinfo{person}{Chin-Wei Chen}, \bibinfo{person}{Ching-Hung Su}, \bibinfo{person}{Kun-Wei Lee}, {and} \bibinfo{person}{Ping-Hao Bair}.} \bibinfo{year}{2020}\natexlab{}.
\newblock \showarticletitle{Malware family classification using active learning by learning}. In \bibinfo{booktitle}{\emph{ICACT}}.
\newblock


\bibitem[Chiew et~al\mbox{.}(2019)]%
        {chiew2019new}
\bibfield{author}{\bibinfo{person}{Kang~Leng Chiew}, \bibinfo{person}{Choon~Lin Tan}, \bibinfo{person}{KokSheik Wong}, \bibinfo{person}{Kelvin~SC Yong}, {and} \bibinfo{person}{Wei~King Tiong}.} \bibinfo{year}{2019}\natexlab{}.
\newblock \showarticletitle{A new hybrid ensemble feature selection framework for machine learning-based phishing detection system}.
\newblock \bibinfo{journal}{\emph{Inf. Scie.}} (\bibinfo{year}{2019}).
\newblock


\bibitem[de~Carvalho~Bertoli et~al\mbox{.}(2023)]%
        {de2023generalizing}
\bibfield{author}{\bibinfo{person}{Gustavo de Carvalho~Bertoli}, \bibinfo{person}{Louren{\c{c}}o Alves~Pereira Junior}, \bibinfo{person}{Osamu Saotome}, {and} \bibinfo{person}{Aldri~Luiz dos Santos}.} \bibinfo{year}{2023}\natexlab{}.
\newblock \showarticletitle{Generalizing intrusion detection for heterogeneous networks: A stacked-unsupervised federated learning approach}.
\newblock \bibinfo{journal}{\emph{Comp. Secur.}} (\bibinfo{year}{2023}).
\newblock


\bibitem[De~Shon(2019)]%
        {de2019information}
\bibfield{author}{\bibinfo{person}{Markus De~Shon}.} \bibinfo{year}{2019}\natexlab{}.
\newblock \showarticletitle{Information Security Analysis as Data Fusion}. In \bibinfo{booktitle}{\emph{FUSION}}.
\newblock


\bibitem[Dias et~al\mbox{.}(2020)]%
        {dias2020go}
\bibfield{author}{\bibinfo{person}{Luis Dias}, \bibinfo{person}{Sim{\~a}o Valente}, {and} \bibinfo{person}{Miguel Correia}.} \bibinfo{year}{2020}\natexlab{}.
\newblock \showarticletitle{Go with the flow: Clustering dynamically-defined netflow features for network intrusion detection with DynIDS}. In \bibinfo{booktitle}{\emph{IEEE NCA}}.
\newblock


\bibitem[Engelen et~al\mbox{.}(2021)]%
        {engelen2021troubleshooting}
\bibfield{author}{\bibinfo{person}{Gints Engelen}, \bibinfo{person}{Vera Rimmer}, {and} \bibinfo{person}{Wouter Joosen}.} \bibinfo{year}{2021}\natexlab{}.
\newblock \showarticletitle{Troubleshooting an intrusion detection dataset: the CICIDS2017 case study}. In \bibinfo{booktitle}{\emph{IEEE S\&P Workshops}}.
\newblock


\bibitem[Fredriksson et~al\mbox{.}(2020)]%
        {fredriksson2020data}
\bibfield{author}{\bibinfo{person}{Teodor Fredriksson}, \bibinfo{person}{David~Issa Mattos}, \bibinfo{person}{Jan Bosch}, {and} \bibinfo{person}{Helena~Holmstr{\"o}m Olsson}.} \bibinfo{year}{2020}\natexlab{}.
\newblock \showarticletitle{Data labeling: An empirical investigation into industrial challenges and mitigation strategies}. In \bibinfo{booktitle}{\emph{PROFES 2020}}. Springer, \bibinfo{pages}{202--216}.
\newblock


\bibitem[Garcia et~al\mbox{.}(2014)]%
        {garcia2014empirical}
\bibfield{author}{\bibinfo{person}{Sebastian Garcia}, \bibinfo{person}{Martin Grill}, \bibinfo{person}{Jan Stiborek}, {and} \bibinfo{person}{Alejandro Zunino}.} \bibinfo{year}{2014}\natexlab{}.
\newblock \showarticletitle{An empirical comparison of botnet detection methods}.
\newblock \bibinfo{journal}{\emph{Comp. Secur.}} (\bibinfo{year}{2014}).
\newblock


\bibitem[Ghafir et~al\mbox{.}(2018)]%
        {ghafir2018detection}
\bibfield{author}{\bibinfo{person}{Ibrahim Ghafir}, \bibinfo{person}{Mohammad Hammoudeh}, \bibinfo{person}{Vaclav Prenosil}, \bibinfo{person}{Liangxiu Han}, \bibinfo{person}{Robert Hegarty}, \bibinfo{person}{Khaled Rabie}, {and} \bibinfo{person}{Francisco~J Aparicio-Navarro}.} \bibinfo{year}{2018}\natexlab{}.
\newblock \showarticletitle{Detection of advanced persistent threat using machine-learning correlation analysis}.
\newblock \bibinfo{journal}{\emph{FGCS}} (\bibinfo{year}{2018}).
\newblock


\bibitem[G{\"o}rnitz et~al\mbox{.}(2009)]%
        {gornitz2009active}
\bibfield{author}{\bibinfo{person}{Nico G{\"o}rnitz}, \bibinfo{person}{Marius Kloft}, \bibinfo{person}{Konrad Rieck}, {and} \bibinfo{person}{Ulf Brefeld}.} \bibinfo{year}{2009}\natexlab{}.
\newblock \showarticletitle{Active learning for network intrusion detection}. In \bibinfo{booktitle}{\emph{AISec}}. \bibinfo{pages}{47--54}.
\newblock


\bibitem[Guerra et~al\mbox{.}(2022)]%
        {guerra2022datasets}
\bibfield{author}{\bibinfo{person}{Jorge~Luis Guerra}, \bibinfo{person}{Carlos Catania}, {and} \bibinfo{person}{Eduardo Veas}.} \bibinfo{year}{2022}\natexlab{}.
\newblock \showarticletitle{Datasets are not enough: Challenges in labeling network traffic}.
\newblock \bibinfo{journal}{\emph{Computers \& Security}} (\bibinfo{year}{2022}).
\newblock


\bibitem[Hannah(2005)]%
        {hannah2005should}
\bibfield{author}{\bibinfo{person}{David~R Hannah}.} \bibinfo{year}{2005}\natexlab{}.
\newblock \showarticletitle{Should I keep a secret? The effects of trade secret protection procedures on employees' obligations to protect trade secrets}.
\newblock \bibinfo{journal}{\emph{Organ. Sci.}} (\bibinfo{year}{2005}).
\newblock


\bibitem[Hong et~al\mbox{.}(2020)]%
        {hong2020phishing}
\bibfield{author}{\bibinfo{person}{Jiwon Hong}, \bibinfo{person}{Taeri Kim}, \bibinfo{person}{Jing Liu}, \bibinfo{person}{Noseong Park}, {and} \bibinfo{person}{Sang-Wook Kim}.} \bibinfo{year}{2020}\natexlab{}.
\newblock \showarticletitle{Phishing url detection with lexical features and blacklisted domains}.
\newblock \bibinfo{journal}{\emph{Adaptive autonomous secure cyber systems}} (\bibinfo{year}{2020}), \bibinfo{pages}{253--267}.
\newblock


\bibitem[Irolla and Dey(2018)]%
        {irolla2018duplication}
\bibfield{author}{\bibinfo{person}{Paul Irolla} {and} \bibinfo{person}{Alexandre Dey}.} \bibinfo{year}{2018}\natexlab{}.
\newblock \showarticletitle{The duplication issue within the drebin dataset}.
\newblock \bibinfo{journal}{\emph{J. Comp. Vir. Hack. Tech.}} (\bibinfo{year}{2018}).
\newblock


\bibitem[Joyce et~al\mbox{.}(2021)]%
        {joyce2021framework}
\bibfield{author}{\bibinfo{person}{Robert~J Joyce}, \bibinfo{person}{Edward Raff}, {and} \bibinfo{person}{Charles Nicholas}.} \bibinfo{year}{2021}\natexlab{}.
\newblock \showarticletitle{A framework for cluster and classifier evaluation in the absence of reference labels}. In \bibinfo{booktitle}{\emph{AISec}}.
\newblock


\bibitem[Kaloudi and Li(2020)]%
        {kaloudi2020ai}
\bibfield{author}{\bibinfo{person}{Nektaria Kaloudi} {and} \bibinfo{person}{Jingyue Li}.} \bibinfo{year}{2020}\natexlab{}.
\newblock \showarticletitle{The ai-based cyber threat landscape: A survey}.
\newblock \bibinfo{journal}{\emph{ACM Computing Surveys (CSUR)}} \bibinfo{volume}{53}, \bibinfo{number}{1} (\bibinfo{year}{2020}), \bibinfo{pages}{1--34}.
\newblock


\bibitem[Kan et~al\mbox{.}(2021)]%
        {kan2021investigating}
\bibfield{author}{\bibinfo{person}{Zeliang Kan}, \bibinfo{person}{Feargus Pendlebury}, \bibinfo{person}{Fabio Pierazzi}, {and} \bibinfo{person}{Lorenzo Cavallaro}.} \bibinfo{year}{2021}\natexlab{}.
\newblock \showarticletitle{Investigating labelless drift adaptation for malware detection}. In \bibinfo{booktitle}{\emph{AISec}}.
\newblock


\bibitem[Kaur et~al\mbox{.}(2022)]%
        {kaur2022trustworthy}
\bibfield{author}{\bibinfo{person}{Davinder Kaur}, \bibinfo{person}{Suleyman Uslu}, \bibinfo{person}{Kaley~J Rittichier}, {and} \bibinfo{person}{Arjan Durresi}.} \bibinfo{year}{2022}\natexlab{}.
\newblock \showarticletitle{Trustworthy artificial intelligence: a review}.
\newblock \bibinfo{journal}{\emph{ACM Computing Surveys (CSUR)}} (\bibinfo{year}{2022}).
\newblock


\bibitem[Koh et~al\mbox{.}(2024)]%
        {koh2024voices}
\bibfield{author}{\bibinfo{person}{Fiona Koh}, \bibinfo{person}{Kathrin Grosse}, {and} \bibinfo{person}{Giovanni Apruzzese}.} \bibinfo{year}{2024}\natexlab{}.
\newblock \showarticletitle{{Voices from the Frontline: Revealing the AI Practitioners’ viewpoint on the EU AI Act}}. In \bibinfo{booktitle}{\emph{HICSS}}.
\newblock


\bibitem[Lemay et~al\mbox{.}(2018)]%
        {lemay2018survey}
\bibfield{author}{\bibinfo{person}{Antoine Lemay}, \bibinfo{person}{Joan Calvet}, \bibinfo{person}{Fran{\c{c}}ois Menet}, {and} \bibinfo{person}{Jos{\'e}~M Fernandez}.} \bibinfo{year}{2018}\natexlab{}.
\newblock \showarticletitle{Survey of publicly available reports on advanced persistent threat actors}.
\newblock \bibinfo{journal}{\emph{Comp. Secur.}} (\bibinfo{year}{2018}).
\newblock


\bibitem[Li and Wang(2017)]%
        {li2017phishbox}
\bibfield{author}{\bibinfo{person}{Jhen-Hao Li} {and} \bibinfo{person}{Sheng-De Wang}.} \bibinfo{year}{2017}\natexlab{}.
\newblock \showarticletitle{PhishBox: An approach for phishing validation and detection}. In \bibinfo{booktitle}{\emph{IEEE DASC/PiCom/DataCom/CyberSciTech}}.
\newblock


\bibitem[Liu et~al\mbox{.}(2021)]%
        {liu2021self}
\bibfield{author}{\bibinfo{person}{Xiao Liu}, \bibinfo{person}{Fanjin Zhang}, \bibinfo{person}{Zhenyu Hou}, \bibinfo{person}{Li Mian}, \bibinfo{person}{Zhaoyu Wang}, \bibinfo{person}{Jing Zhang}, {and} \bibinfo{person}{Jie Tang}.} \bibinfo{year}{2021}\natexlab{}.
\newblock \showarticletitle{Self-supervised learning: Generative or contrastive}.
\newblock \bibinfo{journal}{\emph{IEEE Transactions on knowledge and data engineering}} \bibinfo{volume}{35}, \bibinfo{number}{1} (\bibinfo{year}{2021}), \bibinfo{pages}{857--876}.
\newblock


\bibitem[Luccioni and Rolnick(2023)]%
        {luccioni2023bugs}
\bibfield{author}{\bibinfo{person}{Alexandra~Sasha Luccioni} {and} \bibinfo{person}{David Rolnick}.} \bibinfo{year}{2023}\natexlab{}.
\newblock \showarticletitle{Bugs in the data: How ImageNet misrepresents biodiversity}. In \bibinfo{booktitle}{\emph{AAAI Conference on Artificial Intelligence}}.
\newblock


\bibitem[Mahdavifar and Ghorbani(2019)]%
        {mahdavifar2019application}
\bibfield{author}{\bibinfo{person}{Samaneh Mahdavifar} {and} \bibinfo{person}{Ali~A Ghorbani}.} \bibinfo{year}{2019}\natexlab{}.
\newblock \showarticletitle{Application of deep learning to cybersecurity: A survey}.
\newblock \bibinfo{journal}{\emph{Neurocomputing}} (\bibinfo{year}{2019}).
\newblock


\bibitem[Matsuura et~al\mbox{.}(2021)]%
        {matsuura2021careless}
\bibfield{author}{\bibinfo{person}{Tenga Matsuura}, \bibinfo{person}{Ayako~A Hasegawa}, \bibinfo{person}{Mitsuaki Akiyama}, {and} \bibinfo{person}{Tatsuya Mori}.} \bibinfo{year}{2021}\natexlab{}.
\newblock \showarticletitle{Careless participants are essential for our phishing study: Understanding the impact of screening methods}. In \bibinfo{booktitle}{\emph{Proceedings of the 2021 EuroUSEC}}. \bibinfo{pages}{36--47}.
\newblock


\bibitem[Meyer and Apruzzese(2022)]%
        {meyer2022smartgrid}
\bibfield{author}{\bibinfo{person}{Jacqueline Meyer} {and} \bibinfo{person}{Giovanni Apruzzese}.} \bibinfo{year}{2022}\natexlab{}.
\newblock \showarticletitle{{Cybersecurity in the Smart Grid: Practitioners' Perspective}}. In \bibinfo{booktitle}{\emph{ICSS Workshop (co-located with ACSAC)}}.
\newblock


\bibitem[Miller et~al\mbox{.}(2016)]%
        {miller2016reviewer}
\bibfield{author}{\bibinfo{person}{Brad Miller}, \bibinfo{person}{Alex Kantchelian}, \bibinfo{person}{Michael~Carl Tschantz}, \bibinfo{person}{Sadia Afroz}, \bibinfo{person}{Rekha Bachwani}, \bibinfo{person}{Riyaz Faizullabhoy}, \bibinfo{person}{Ling Huang}, \bibinfo{person}{Vaishaal Shankar}, \bibinfo{person}{Tony Wu}, \bibinfo{person}{George Yiu}, {et~al\mbox{.}}} \bibinfo{year}{2016}\natexlab{}.
\newblock \showarticletitle{Reviewer integration and performance measurement for malware detection}. In \bibinfo{booktitle}{\emph{Proc. Int. Conf. DIMVA}}. \bibinfo{pages}{122--141}.
\newblock


\bibitem[Mjolsness and DeCoste(2001)]%
        {mjolsness2001machine}
\bibfield{author}{\bibinfo{person}{Eric Mjolsness} {and} \bibinfo{person}{Dennis DeCoste}.} \bibinfo{year}{2001}\natexlab{}.
\newblock \showarticletitle{Machine learning for science: state of the art and future prospects}.
\newblock \bibinfo{journal}{\emph{Science}} (\bibinfo{year}{2001}).
\newblock


\bibitem[Montaruli et~al\mbox{.}(2023)]%
        {montaruli2023raze}
\bibfield{author}{\bibinfo{person}{Biagio Montaruli}, \bibinfo{person}{Luca Demetrio}, \bibinfo{person}{Maura Pintor}, \bibinfo{person}{Battista Biggio}, \bibinfo{person}{Luca Compagna}, {and} \bibinfo{person}{Davide Balzarotti}.} \bibinfo{year}{2023}\natexlab{}.
\newblock \showarticletitle{Raze to the Ground: Query-Efficient Adversarial HTML Attacks on Machine-Learning Phishing Webpage Detectors}. In \bibinfo{booktitle}{\emph{AISec}}.
\newblock


\bibitem[Mu{\~n}oz-Gonz{\'a}lez et~al\mbox{.}(2019)]%
        {munoz2019challenges}
\bibfield{author}{\bibinfo{person}{Luis Mu{\~n}oz-Gonz{\'a}lez}, \bibinfo{person}{Javier Carnerero-Cano}, \bibinfo{person}{Kenneth~T Co}, {and} \bibinfo{person}{Emil~C Lupu}.} \bibinfo{year}{2019}\natexlab{}.
\newblock \showarticletitle{Challenges and Advances in Adversarial Machine Learning}.
\newblock \bibinfo{journal}{\emph{Resilience and Hybrid Threats}} (\bibinfo{year}{2019}), \bibinfo{pages}{102--120}.
\newblock


\bibitem[Nisioti et~al\mbox{.}(2018)]%
        {nisioti2018intrusion}
\bibfield{author}{\bibinfo{person}{Antonia Nisioti}, \bibinfo{person}{Alexios Mylonas}, \bibinfo{person}{Paul~D Yoo}, {and} \bibinfo{person}{Vasilios Katos}.} \bibinfo{year}{2018}\natexlab{}.
\newblock \showarticletitle{From intrusion detection to attacker attribution: A comprehensive survey of unsupervised methods}.
\newblock \bibinfo{journal}{\emph{IEEE Communications Surveys \& Tutorials}} \bibinfo{volume}{20}, \bibinfo{number}{4} (\bibinfo{year}{2018}), \bibinfo{pages}{3369--3388}.
\newblock


\bibitem[Pape et~al\mbox{.}(2023)]%
        {pape2023limitations}
\bibfield{author}{\bibinfo{person}{David Pape}, \bibinfo{person}{Sina D{\"a}ubener}, \bibinfo{person}{Thorsten Eisenhofer}, \bibinfo{person}{Antonio~Emanuele Cin{\`a}}, {and} \bibinfo{person}{Lea Sch{\"o}nherr}.} \bibinfo{year}{2023}\natexlab{}.
\newblock \showarticletitle{On the Limitations of Model Stealing with Uncertainty Quantification Models}.
\newblock \bibinfo{journal}{\emph{ESANN}}.
\newblock


\bibitem[Pendlebury et~al\mbox{.}(2019)]%
        {pendlebury2019tesseract}
\bibfield{author}{\bibinfo{person}{Feargus Pendlebury}, \bibinfo{person}{Fabio Pierazzi}, \bibinfo{person}{Roberto Jordaney}, \bibinfo{person}{Johannes Kinder}, {and} \bibinfo{person}{Lorenzo Cavallaro}.} \bibinfo{year}{2019}\natexlab{}.
\newblock \showarticletitle{$\{$TESSERACT$\}$: Eliminating experimental bias in malware classification across space and time}. In \bibinfo{booktitle}{\emph{USENIX Security 19}}. \bibinfo{pages}{729--746}.
\newblock


\bibitem[Rashidi et~al\mbox{.}(2017)]%
        {rashidi2017android}
\bibfield{author}{\bibinfo{person}{Bahman Rashidi}, \bibinfo{person}{Carol Fung}, {and} \bibinfo{person}{Elisa Bertino}.} \bibinfo{year}{2017}\natexlab{}.
\newblock \showarticletitle{Android malicious application detection using support vector machine and active learning}. In \bibinfo{booktitle}{\emph{CNSM}}.
\newblock


\bibitem[Ren et~al\mbox{.}(2021)]%
        {ren2021survey}
\bibfield{author}{\bibinfo{person}{Pengzhen Ren}, \bibinfo{person}{Yun Xiao}, \bibinfo{person}{Xiaojun Chang}, \bibinfo{person}{Po-Yao Huang}, \bibinfo{person}{Zhihui Li}, \bibinfo{person}{Brij~B Gupta}, \bibinfo{person}{Xiaojiang Chen}, {and} \bibinfo{person}{Xin Wang}.} \bibinfo{year}{2021}\natexlab{}.
\newblock \showarticletitle{A survey of deep active learning}.
\newblock \bibinfo{journal}{\emph{ACM computing surveys (CSUR)}} \bibinfo{volume}{54}, \bibinfo{number}{9} (\bibinfo{year}{2021}), \bibinfo{pages}{1--40}.
\newblock


\bibitem[Robertson et~al\mbox{.}(2006)]%
        {robertson2006using}
\bibfield{author}{\bibinfo{person}{William Robertson}, \bibinfo{person}{Giovanni Vigna}, \bibinfo{person}{Christopher Kruegel}, \bibinfo{person}{Richard~A Kemmerer}, {et~al\mbox{.}}} \bibinfo{year}{2006}\natexlab{}.
\newblock \showarticletitle{Using generalization and characterization techniques in the anomaly-based detection of web attacks}. In \bibinfo{booktitle}{\emph{NDSS}}.
\newblock


\bibitem[Sarker et~al\mbox{.}(2020)]%
        {sarker2020cybersecurity}
\bibfield{author}{\bibinfo{person}{Iqbal~H Sarker}, \bibinfo{person}{ASM Kayes}, \bibinfo{person}{Shahriar Badsha}, \bibinfo{person}{Hamed Alqahtani}, \bibinfo{person}{Paul Watters}, {and} \bibinfo{person}{Alex Ng}.} \bibinfo{year}{2020}\natexlab{}.
\newblock \showarticletitle{Cybersecurity data science: an overview from machine learning perspective}.
\newblock \bibinfo{journal}{\emph{Journal of Big data}}  \bibinfo{volume}{7} (\bibinfo{year}{2020}), \bibinfo{pages}{1--29}.
\newblock


\bibitem[Shapira et~al\mbox{.}(2023)]%
        {shapira2023phantom}
\bibfield{author}{\bibinfo{person}{Avishag Shapira}, \bibinfo{person}{Alon Zolfi}, \bibinfo{person}{Luca Demetrio}, \bibinfo{person}{Battista Biggio}, {and} \bibinfo{person}{Asaf Shabtai}.} \bibinfo{year}{2023}\natexlab{}.
\newblock \showarticletitle{Phantom Sponges: Exploiting Non-Maximum Suppression to Attack Deep Object Detectors}. In \bibinfo{booktitle}{\emph{IEEE/CVF Winter Conf. Appl. Comp. Vision}}.
\newblock


\bibitem[Sharafaldin et~al\mbox{.}(2018)]%
        {sharafaldin2018toward}
\bibfield{author}{\bibinfo{person}{Iman Sharafaldin}, \bibinfo{person}{Habibi Lashkari}, {and} \bibinfo{person}{Ali~A Ghorbani}.} \bibinfo{year}{2018}\natexlab{}.
\newblock \showarticletitle{Toward generating a new intrusion detection dataset and intrusion traffic characterization.}
\newblock \bibinfo{journal}{\emph{ICISSp}} (\bibinfo{year}{2018}).
\newblock


\bibitem[Silva et~al\mbox{.}(2023)]%
        {silva2023self}
\bibfield{author}{\bibinfo{person}{Thalles Silva}, \bibinfo{person}{Helio Pedrini}, {and} \bibinfo{person}{Ad{\'\i}n~Ram{\'\i}rez Rivera}.} \bibinfo{year}{2023}\natexlab{}.
\newblock \showarticletitle{Self-supervised Learning of Contextualized Local Visual Embeddings}. In \bibinfo{booktitle}{\emph{VIPriors 4}}.
\newblock


\bibitem[Sommer and Paxson(2010)]%
        {sommer2010outside}
\bibfield{author}{\bibinfo{person}{Robin Sommer} {and} \bibinfo{person}{Vern Paxson}.} \bibinfo{year}{2010}\natexlab{}.
\newblock \showarticletitle{Outside the closed world: On using machine learning for network intrusion detection}. In \bibinfo{booktitle}{\emph{IEEE S\&P}}.
\newblock


\bibitem[Tian et~al\mbox{.}(2018)]%
        {tian2018needle}
\bibfield{author}{\bibinfo{person}{Ke Tian}, \bibinfo{person}{Steve~TK Jan}, \bibinfo{person}{Hang Hu}, \bibinfo{person}{Danfeng Yao}, {and} \bibinfo{person}{Gang Wang}.} \bibinfo{year}{2018}\natexlab{}.
\newblock \showarticletitle{Needle in a haystack: Tracking down elite phishing domains in the wild}. In \bibinfo{booktitle}{\emph{IMC}}.
\newblock


\bibitem[Van~Ede et~al\mbox{.}(2022)]%
        {van2022deepcase}
\bibfield{author}{\bibinfo{person}{Thijs Van~Ede}, \bibinfo{person}{Hojjat Aghakhani}, \bibinfo{person}{Noah Spahn}, \bibinfo{person}{Riccardo Bortolameotti}, \bibinfo{person}{Marco Cova}, \bibinfo{person}{Andrea Continella}, \bibinfo{person}{Maarten van Steen}, \bibinfo{person}{Andreas Peter}, \bibinfo{person}{Christopher Kruegel}, {and} \bibinfo{person}{Giovanni Vigna}.} \bibinfo{year}{2022}\natexlab{}.
\newblock \showarticletitle{Deepcase: Semi-supervised contextual analysis of security events}. In \bibinfo{booktitle}{\emph{2022 IEEE SP}}. \bibinfo{pages}{522--539}.
\newblock


\bibitem[Weller et~al\mbox{.}(2018)]%
        {weller2018open}
\bibfield{author}{\bibinfo{person}{Susan~C Weller}, \bibinfo{person}{Ben Vickers}, \bibinfo{person}{H~Russell Bernard}, \bibinfo{person}{Alyssa~M Blackburn}, \bibinfo{person}{Stephen Borgatti}, \bibinfo{person}{Clarence~C Gravlee}, {and} \bibinfo{person}{Jeffrey~C Johnson}.} \bibinfo{year}{2018}\natexlab{}.
\newblock \showarticletitle{Open-ended interview questions and saturation}.
\newblock \bibinfo{journal}{\emph{PloS one}} (\bibinfo{year}{2018}).
\newblock


\bibitem[Xu et~al\mbox{.}(2020)]%
        {xu2020method}
\bibfield{author}{\bibinfo{person}{Congyuan Xu}, \bibinfo{person}{Jizhong Shen}, {and} \bibinfo{person}{Xin Du}.} \bibinfo{year}{2020}\natexlab{}.
\newblock \showarticletitle{A method of few-shot network intrusion detection based on meta-learning framework}.
\newblock \bibinfo{journal}{\emph{IEEE TIFS}} (\bibinfo{year}{2020}).
\newblock


\bibitem[Yang et~al\mbox{.}(2017)]%
        {yang2017multi}
\bibfield{author}{\bibinfo{person}{Jun Yang}, \bibinfo{person}{Pengpeng Yang}, \bibinfo{person}{Xiaohui Jin}, {and} \bibinfo{person}{Qian Ma}.} \bibinfo{year}{2017}\natexlab{}.
\newblock \showarticletitle{Multi-classification for malicious URL based on improved semi-supervised algorithm}. In \bibinfo{booktitle}{\emph{IEEE CSE}}.
\newblock


\bibitem[Zhang et~al\mbox{.}(2021)]%
        {zhang2021network}
\bibfield{author}{\bibinfo{person}{Yong Zhang}, \bibinfo{person}{Jie Niu}, \bibinfo{person}{Guojian He}, \bibinfo{person}{Lin Zhu}, {and} \bibinfo{person}{Da Guo}.} \bibinfo{year}{2021}\natexlab{}.
\newblock \showarticletitle{Network Intrusion Detection Based on Active Semi-supervised Learning}. In \bibinfo{booktitle}{\emph{DSN-W}}.
\newblock


\bibitem[Zhu and Yang(2019)]%
        {zhu2019tripartite}
\bibfield{author}{\bibinfo{person}{Yanqiao Zhu} {and} \bibinfo{person}{Kai Yang}.} \bibinfo{year}{2019}\natexlab{}.
\newblock \showarticletitle{Tripartite active learning for interactive anomaly discovery}.
\newblock \bibinfo{journal}{\emph{IEEE Access}}  \bibinfo{volume}{7} (\bibinfo{year}{2019}), \bibinfo{pages}{63195--63203}.
\newblock


\end{thebibliography}
}

\appendix
\section{Background on Active Learning}
\label{app:active}
\noindent
The fundamental principle of \textit{active learning} (AL) is to optimize the labeling procedure by ``suggesting'' to a given (human) annotator which (unlabelled) samples should be provided with their ground truth. The intuition is that some samples are ``more informative'' than others: by having an ML model be trained on such samples, it is possible to improve its learning in a cost-effective way~\cite{apruzzese2022sok}.

Formally, given an ML model \smamath{M_0} (having performance \smamath{\mu_0}) and an unlabelled dataset \smabb{U}, AL methods seek to identify which samples \smamath{x_a}\smamath{\in}\smabb{U}, when used to update \smamath{M_0} (after being correctly labeled), yield an ML model \smamath{M_a} whose performance \smamath{\mu_a} is superior to the performance \smamath{\bar{\mu}} of another ML model \smamath{\bar{M}} obtained by updating the original ML model \smamath{M_0} with any other sample \smamath{\bar{x}}\smamath{\in}\smabb{U} (with \smamath{\bar{x}}\smamath{\neq}\smamath{x_a}).

Among the many methods~\cite{ren2021survey} encompassed by AL, a popular one is \textbf{uncertainty sampling}~\cite{rashidi2017android}, which leverages the predictions of a pre-trained ML model \smamath{M_0} as a guide for the ``suggestions''. The idea is that \smamath{M_0} is likely to learn `more' from samples that it cannot properly recognize.\footnote{In a sense, this is the principle of adversarial training~\cite{munoz2019challenges}.} Hence, by computing the ``uncertainty'' (e.g.,~\cite{pape2023limitations}) of \smamath{M_0} on the samples \smamath{x}\smamath{\in}\smabb{U}, it is possible to optimize the labeling by having the annotator focus only on those samples that have the highest uncertainty by \smamath{M_0}. Such a procedure has been shown to be significantly more efficient than random sampling~\cite{apruzzese2022sok}.

From a research perspective, it is possible to simulate the abovementioned workflow as follows. First, given a (labeled) dataset \smabb{D}, the researcher must reserve a subset \smabb{E} used for performance evaluation; and isolate a (small) portion which is considered to be labeled (i.e., \smabb{L}), and then consider the remaining samples\footnote{Importantly, $\mathbb{E}\cap(\mathbb{L}\cup\mathbb{U})=\varnothing$} as unlabelled (i.e., \smabb{U)}. Next, the researcher must train an ML model \smamath{M_0} on \smabb{L}, compute its performance \smamath{\mu_0} on \smabb{E}, and use \smamath{M_0} to analyze the samples in \smabb{U}, ensuring to store the confidence (or uncertainty) of each prediction (which can be discarded). Then, the researcher must order the resulting samples according to their confidence, and use the given labeling budget \smacal{B} to `move' the samples with the lowest confidence (or highest uncertainty) from \smabb{U} to \smabb{L} (thereby obtaining \smamath{\mathbb{L}_a}), but \textit{by assigning them with the correct label} (which the researcher knows). Finally, the researcher must re-train \smamath{M_0} on \smamath{\mathbb{L}_a} (obtaining \smamath{M_a}), and assess the resulting performance \smamath{\mu_a} on \smabb{E}. Ideally, \smamath{\mu_a} should be largely superior to \smamath{\mu_0}, and superior to the \smamath{\bar{\mu}} of any \smamath{\bar{M}} yielded by re-training \smamath{M_0} on any updated version of \smabb{L} obtained by using the same budget \smacal{B} through random sampling from \smabb{U}. This process can be repeated many times, each time taking some samples from \smabb{U} and moving them to \smabb{L}.

\section{Runtime}
\label{app:runtime}

\noindent
We perform our experiments on an Intel i9-12900H CPU (6 cores @ 5GHz) with 64GB of RAM. The runtime for performing all the experiments in §\ref{ssec:baseline} was 666 seconds; for §\ref{ssec:mislabeling}, it was 146s; for §\ref{ssec:active}, it was 8445s. More details are in our repository~\cite{ourRepo}.

\end{document}